\documentclass[twocolumn]{aastex61}
\usepackage{colortbl}
\received{August 1, 2017}
\revised{October 17, 2017}
\accepted{October 18,2017}
\submitjournal{ApJS}

\shorttitle{LSST and Euclid}
\shortauthors{Rhodes et al.}

\begin{document}

\title{Scientific Synergy  between LSST and Euclid}

\correspondingauthor{Jason Rhodes}
\email{jason.d.rhodes@jpl.nasa.gov}

\def\caltech{California Institute of Technology, 1201 East California Blvd., Pasadena, CA 91125, USA}
\def\ipac{IPAC, California Institute of Technology, 1200 East California Blvd., Pasadena, CA 91125, USA}
\def\cmu{McWilliams Center for Cosmology, Department of Physics, Carnegie Mellon University, Pittsburgh, PA 15213, USA}
\def\ctio{Cerro Tololo Inter-American Observatory, La Serena, Chile}
\def\jpl{Jet Propulsion Laboratory, California Institute of Technology, Pasadena, CA 91109, USA}
\def\lcogt{LCOGT, University of California, Santa Barbara, CA \ldots}
\def\princeton{Department of Astrophysical Sciences, Princeton University, Princeton, NJ 08544, USA}
\def\scsu{Southern Connecticut State University, \ldots}
\def\slac{SLAC National Accelerator Laboratory, 2575 Sand Hill Road, MS29, Menlo Park, CA 94025, USA}
\def\stanford{Physics Department, Stanford University, Stanford, CA, 94305, USA}
\def\uw{Department of Astronomy, University of Washington, Box 351580, U.W., Seattle WA 98195, USA}
\def\mssl{Mullard Space Science Laboratory, University College London, Surrey, RH5 6NT, UK}
\def\cornell{Department of Astronomy, Cornell University, Ithaca, NY 14853, USA}
\def\bonn{Argelander-Institut f\"{u}r Astronomie, Universit\"{a}t Bonn, Auf dem H\"{u}gel 71, 53121, Bonn, Germany}
\def\cea{CEA Saclay, \ldots}
\def\portsmouth{Institute of Cosmology \& Gravitation (ICG),
University of Portsmouth,
Dennis Sciama Building
Portsmouth, PO1 3FX, UK}
\def\ucsc{Department of Astronomy and Astrophysics, University of California, Santa Cruz, 1156 High Street, Santa Cruz, CA 95064, USA }
\def\durham{Durham University, Centre for Extragalactic Astrophysics, South Road, Durham DH1 3LE, UK}
\def\edi{Institute for Astronomy, School of Physics and Astronomy, Royal Observatory Edinburgh,
   Blackford Hill, Edinburgh, EH9 3HJ, UK}
\def\pitt{Department of Physics and Astronomy and PITT PACC, University of Pittsburgh, 3941 O’Hara St., Pittsburgh, PA 15260, USA}
\def\stony{Department of Physics and Astronomy, Stony Brook University, Stony Brook, NY 11794, USA}
\def\nottingham{School of Physics and Astronomy, University of Nottingham, University Park, Nottingham NG7 2RD, UK}
\def\birmingham{School of Physics and Astronomy, University of Birmingham, Birmingham, B15 2TT, UK}
\def\toronto{Dunlap Institute for Astronomy and Astrophysics \& the Department of Astronomy and Astrophysics, University of Toronto, 50 St. George Street, Toronto, Ontario, Canada M5S 3H4}
\def\bristol{HH Wills Physics Laboratory, University of Bristol, Tyndall Avenue, Bristol, BS8 1TL, UK}
\def\inaf{I.N.A.F. - Osservatorio Astronomico di Roma Via Frascati 33 - 00040 Monte Porzio Catone (Roma), Italy}

\def\iap{Institut d'Astrophysique de Paris, UMR7095 CNRS, Universit\'e Pierre \& Marie Curie, 98 bis boulevard Arago, 75014 Paris, France}
\def\irfu{IRFU, Service d'Astrophysique, CEA Saclay, F-91191 Gif-sur-Yvette Cedex, France}
\def\france{AstroParticule et Cosmologie, Universit\'e Paris Diderot, CNRS, CEA, Observatoire de Paris, Sorbonne Paris Cit\'e B\^atiment Condorcet, 10, rue Alice Domon et L\'eonie Duquet, F-75205 Paris Cedex 13, France }
\def\in2p3{Centre de Calcul de l’IN2P3, USR 6402 du CNRS-IN2P3, 43 Bd. du 11 Novembre 1918, 69622 Villeurbanne Cedex, France}
\def\apc{APC, Astroparticule et Cosmologie, Universit\'e Paris Diderot, CNRS/IN2P3, CEA/Irfu, Observatoire de Paris, Sorbonne Paris Cit\'e, 10, rue Alice Domon \& L\'eonie Duquet, 75205 Paris Cedex 13, France}
\def\aim{Laboratoire AIM, CEA/IRFU/Service d'Astrophysique, CNRS, Universit\'e Paris Diderot, B\^at. 709, 91191 Gif-sur-Yvette, France}
\def\cnrs{Universit\'e C{\^o}te d'Azur, Observatoire de la C{\^o}te d'Azur, CNRS, Laboratoire Lagrange, France}
\def\lapp{Laboratoire d'Annecy de Physique des Particules (LAPP), Universit\'e Savoie Mont Blanc, CNRS/IN2P3, F-74941 Annecy, France}
\def\paris{Observatoire de Paris, PSL Research University, 61, avenue de l'Observatoire, F-75014 Paris, France}
\def\upd{Universit\'e Paris Diderot, AIM, Sorbonne Paris Cit\'e, CEA, CNRS, F-91191 Gif-sur-Yvette Cedex, France}

\author{Jason Rhodes}
\affil{\jpl}\affil{\caltech}

\author{Robert C. Nichol}
\affil{\portsmouth}
\author{\'Eric Aubourg}
\affil{\apc}
\author{Rachel Bean}
\affil{\cornell}
\author{Dominique Boutigny},
\affil{\lapp}
\author{Malcolm N. Bremer}
\affil{\bristol}
\author{Peter Capak}
\affil{\ipac}
\author{Vincenzo Cardone}
\affil{\inaf}
\author{Beno\^it Carry}
\affil{\cnrs}
\author{Christopher J. Conselice}
\affil{\nottingham}
\author{Andrew J. Connolly}
\affil{\uw}
\author{Jean-Charles Cuillandre}
\affil{\aim}\affil{\paris}
\author{N. A. Hatch}
\affil{\nottingham}
\author {George Helou}
\affil{\ipac}
\author{Shoubaneh Hemmati}
\affil{\ipac}
\author{Hendrik Hildebrandt}
\affil{\bonn}
\author{Ren\'{e}e Hlo\v{z}ek}
\affil{\toronto}
\author{Lynne Jones}
\affil{\uw}
\author{Steven Kahn}
\affil{\slac}
\author{Alina Kiessling}
\affil{\jpl}
\author{Thomas Kitching}
\affil{\mssl}
\author{Robert Lupton}
\affil{\princeton}
\author{Rachel Mandelbaum}
\affil{\cmu}
\author{Katarina Markovic}
\affil{\portsmouth}
\author{Phil Marshall}
\affil{\slac}
\author{Richard Massey}
\affil{\durham}
\author{Ben J. Maughan}
\affil{\bristol}
\author{Peter Melchior}
\affil{\princeton}
\author{Yannick Mellier}
\affil{\iap}\affil{\irfu}
\author{Jeffrey A. Newman}
\affil{\pitt}
\author{Brant Robertson}
\affil{\ucsc}
\author{Marc Sauvage}
\affil{\aim}
\author{Tim Schrabback}
\affil{\bonn}
\author{Graham P. Smith}
\affil{\birmingham}
\author{Michael A. Strauss}
\affil{\princeton}
\author{Andy Taylor}
\affil{\edi}
\author{Anja Von Der Linden}
\affil{\stony}





\begin{abstract}

\emph{Euclid} and the  Large Synoptic Survey Telescope (LSST) are poised to dramatically change the astronomy landscape early in the next decade.  The combination of high cadence, deep, wide-field optical photometry from LSST with high resolution, wide-field optical photometry, and near-infrared photometry and spectroscopy from \emph{Euclid} will be powerful for addressing a wide range of astrophysical questions.  We explore \emph{Euclid}/LSST synergy,   ignoring the political issues associated with  data access to focus  on the scientific, technical, and financial benefits of  coordination.   We focus primarily on dark energy cosmology, but also discuss galaxy evolution, transient objects, solar system science, and galaxy cluster studies. We concentrate on synergies that require coordination in cadence or survey overlap, or would benefit from pixel-level co-processing that is beyond the scope of what is currently planned, rather than scientific programs that could be accomplished only at the catalog level without coordination in data  processing or  survey strategies. We provide two quantitative examples of scientific synergies: the decrease in photo-z errors (benefitting many science cases) when high resolution \emph{Euclid} data are used for LSST photo-z determination, and the resulting increase in weak lensing signal-to-noise ratio from smaller photo-z errors. We briefly discuss other areas of coordination, including high performance computing resources and calibration data. Finally, we address concerns about the loss of independence and potential cross-checks between the two missions and the potential consequences of not collaborating.

\end{abstract}




\section{Introduction}
\label{sec:overview}

We present a broad overview of the potential synergies between the European Space Agency's \emph{Euclid} mission\footnote{\tt http://www.euclid-ec.org} \citep{2011arXiv1110.3193L} and the Large Synoptic Survey Telescope\footnote{\tt http://www.lsst.org} \citep[LSST; ][]{2008arXiv0805.2366I, 2009-Book-LSST}. This work builds on the white paper by \citet{JainEtal2015}, which started the discussion on the scientific and technical merits of combining LSST, {\it Euclid} and the \emph{Wide Field Infrared Survey Telescope} \citep[\emph{WFIRST}; ][]{spergel15}. As this paper is a joint LSST and {\it Euclid} effort, we only focus herein on synergies between these two facilities and continue to emphasize the scientific and technical gains from combining these data. We do not attempt to provide an exhaustive review of all possible gains; we leave that for others to continue to explore as well as to develop more quantitative analyses that are beyond the scope of this paper. We assume that future work will require some ``figures of merit''  to help optimize these gains, especially to help in the allocation of (limited) resources.

We have not attempted to broaden this paper beyond LSST and {\it Euclid} to include other international projects including \emph{WFIRST}, the Dark Energy Spectroscopic Instrument (DESI\footnote{\tt http://desi.lbl.gov/}), or the  Square Kilometre Array (SKA\footnote{\tt http://skatelescope.org/}). Other authors have begun to explore such synergies and we  refer the reader to those other papers \citep[e.g.][]{JainEtal2015,LSST-SKA,Euclid-SKA}.

\subsection{Dark Energy Experiments}

By the late 1990s, there was definitive evidence that the universe was in a phase of accelerated expansion.  The unknown cause of this acceleration was dubbed dark energy. At the start of the millennium, it became clear that more data were required
to determine the nature of dark energy. At the same time, new ways to analyze astronomical data from
large wide-field surveys were being developed, in particular weak gravitational lensing
and Baryon Acoustic Oscillations (BAO) observed via galaxy clustering. Techniques for  classifying supernovae also continued to advance. The diversity of this field at this time has been captured in the
ESO-ESA Report on dark energy \citep{2006ewg} and the Dark Energy Task Force (DETF) reports \citep{DETF, DETF2}.

What evolved from this intense period of discussion of new experiments was an international program
that, over the  two decades from 2010 to 2030, will observe successively larger areas of the sky
at wavelengths from the optical to infrared (and eventually, radio), from both the ground and space. It was realized early on that future dark energy experiments should exploit the combination of multiple cosmological techniques as this would lead to more secure dark energy
measurements, tighter control of systematics,  and the ability to distinguish between dark energy and possible modifications to general relativity (modified gravity). Thus, both \emph{Euclid} and LSST have evolved through a process of niche ,separation and
complementarity to existing proposals to encompass a unique combination of methodologies and
sensitivity to systematic effects.

Each experiment was designed to simultaneously minimize statistical
error and systematics  for  dark energy studies, whilst finding maximum complementarity
with existing, ongoing surveys. The resulting projects approximately follow a ``staged''
classification proposed by the DETF report, whereby early experiments designed
to confirm the existence of dark energy are classified as ``Stage II'', mid-cycle experiments
that can determine cosmological parameters within our current paradigm to high accuracy and test
systematic effects are ``Stage III'', and nearly all-sky experiments that have the capability
to determine the physical nature of dark energy are ``Stage IV''. The majority of Stage II experiments
are now complete, proving the existence of dark energy beyond doubt for a majority of cosmologists \citep{2013PhR...530...87W}.

Stage III experiments are now coming to maturity, with surveys including the Kilo Degree Survey \citep[KiDS]{kids}, the Baryon Oscillation Spectroscopic Survey \citep[BOSS]{BOSS}, the Subaru HyperSuprimeCam Survey \citep[HSC]{2017arXiv170405858A}, and the Dark Energy Survey    \citep[DES]{DESYEAR1} delivering new results. These latest surveys still highlight possible tensions within our present cosmological model including ongoing differences between low and high redshift measurements of the Hubble constant (\citealt{2016arXiv161004606J} and \citealt{2017MNRAS.465.4914B}) and the growth of structure \citep{kids, 2016MNRAS.459..971K, 2017arXiv170700483E}. These differences could be due to unresolved systematic uncertainties that would need to be understood to allow us to make further progress, or to the first signs of our incomplete cosmological understanding that may require new physics.

There are now several Stage IV experiments under construction, including \emph{Euclid}  and LSST. The methods used by these experiments all drive the
experiment design toward large wide-field surveys, ideally with high spatial
resolution and high cadence of observations. Neither \emph{Euclid} or LSST individually satisfy all
of these criteria, but together, they do. The sum of all data from these two facilities will
survey the majority of the extragalactic sky, with high spatial resolution, a
rapid cadence, from ground and space, and in multiple filters from blue optical to near infrared (NIR).
This abundance of new survey data will change astronomy in ways that
go far beyond dark energy and follows the rich history of other
surveys like the Sloan Digital Sky Survey \citep[SDSS,][]{Yorketal2000}
where the access to high-quality public data has promoted a
wealth of legacy research (e.g. thousands of journal papers) and public
engagement opportunities \citep[e.g. GalaxyZoo\footnote{\tt http://www.galaxyzoo.org};][]{2011MNRAS.410..166L,2010AEdRv...9a0103R}.

 \emph{Euclid} and LSST each will survey a significant fraction of the sky starting at the beginning of the next decade. We summarize
each of these facilities below and provide more details about the intended surveys in \S\ref{sec:coverage}:
\begin{itemize}
\item
LSST is a
ground-based telescope
with a primary mirror with an effective diameter of $6.7$m designed to provide a time-domain imaging survey
of the entire southern hemisphere ($\simeq18,000$\,square degrees)
in six optical bands ($ugrizy$) with an average seeing of 0.7 arcsec. First light is scheduled for 2020, with survey operations commencing in 2022 and
running for the following 10 years to gradually accumulate depth to
$r\sim27.5$ magnitudes ($5\sigma$ point source). The science objectives of LSST are multifold, but among the
primary scientific drivers is the elucidation of the nature of dark
energy and dark matter through a combination of probes associated with
statistical analyses of  billions of galaxies and hundreds of thousands of supernovae.
\item
\emph{Euclid }is a 1.2m space-based telescope designed to image $15,000$\
square degrees of sky in one broad optical band (VIS = $r+i+z$ covering approximately 5400 to 9000 \AA)
with pristine image quality (0.16$^{\prime\prime}$ FWHM) to an
expected depth of at least $24.5$ AB magnitudes (10$\sigma$ 1$^{\prime\prime}$~extended
source) and three near-infrared bands covering approximately 0.95 to 2 $\mu$m down to
$Y=24.0$, $J=24.2$ and $H=23.9$ AB magnitudes (5$\sigma$, point source) respectively.
\emph{Euclid} will also obtain near-infrared low-resolution ($R\sim250$) grism
spectroscopy over the same area of the sky using a 
``red'' grism (1.25 to 1.85 $\mu$m). \emph{Euclid} is officially scheduled for launch in late 2020, with science operations of a least 6.5 years.
Although \emph{Euclid} data will also address a range of scientific questions, the primary goal of the
mission is the elucidation of the nature of dark energy (along with possible modifications to general relativity) through investigations of
the distance-redshift relation of galaxies and the evolution of cosmic
structures.
\end{itemize}

With these features, it is clear that \emph{Euclid} and LSST are complementary.
Consequently, as the scientific goals and schedule overlap, it is
natural to ask whether cooperation between the two experiments can lead to both increased efficiency and scientific reach. Indeed, the limitations and
risks faced by one are mitigated by the strengths of the other e.g. LSST observes the sky from the ground, while \emph{Euclid} has a finite lifetime and constrained cadence due to degradation of the CCDs in the space environment and limited propulsion gas needed to point the telescope.

This document aims to outline the benefits of scientific cooperation, mainly at the level of survey coordination, information interchange, and joint data processing, as well as to provide a technical framework for how the coordinated efforts might be arranged.  We also seek to identify potential scientific risks associated with such coordination that result from the loss of independent analyses that would occur due to sharing the data and associated tools (see Appendix \ref{sec:independence} for a discussion of the merits of independence).  This document should inform future discussions between the \emph{Euclid} Consortium (EC), the LSST Project, and associated LSST science collaborations about how to achieve maximal joint cosmological science in the next decade.

The paper is organized as follows:  \S\ref{sec:cosmology} outlines the benefits of possible coordination for cosmology in more detail, and some non-cosmology science and technical cases for coordination are outlined in \S\ref{sec:other_benefits}.  \S\ref{sec:tech_level} describes the technical level of coordination required, and \S\ref{sec:howto} describes several models in which we might achieve that coordination. What we still need in order to achieve the described coordination is described in \S\ref{sec:needs} and we offer concluding remarks in \S\ref{sec:conclusions}.  A discussion of the pros and cons of coordinating two experiments, and a warning about what a lack of coordination might mean, is included in Appendix \ref{sec:appendix}.




\section{Benefits for Cosmology}
\label{sec:cosmology}
In this section, we detail some of the benefits of coordination for cosmological science.  Benefits to non-cosmological science are detailed in \S\ref{sec:noncosmo}.

\subsection{Object Detection}
\label{sec:objectdetection}

The optical images from LSST and \emph{Euclid's} VIS instrument differ in three main characteristics: spatial resolution, number of filters,  and depth.
\emph{Euclid's} Near Infrared Spectrometer and Photometer (NISP) differs from LSST and VIS by covering the near-infrared (NIR) regime, and by being significantly undersampled.
These differences can be exploited to optimize object detection and characterization.

The \emph{Euclid} VIS instrument has a spatial resolution of $0.1\arcsec$ per pixel with a $\leq0.18\arcsec$ width of the point spread function  \citep[PSF;][]{Cropper2016}. This is compared to the anticipated median seeing for LSST of 0.7$\arcsec$.
This difference in resolution can enable the recognition of objects
that would be unresolved by LSST, improve separation of stars from
galaxies, and improve the deblending of objects. This works for objects
that are brighter than $\simeq25$ AB magnitudes, the present design
limit of VIS for a 1$^{\prime\prime}$ extended source.
LSST will have the advantage of a point-source detection depth of $\sim27.0$ AB magnitudes in the $i$-band, which is superior for the detection of fainter objects and extended emission, e.g. low surface brightness galaxies and intracluster light (especially at wavelengths shorter than  the \emph{Euclid} VIS filter band).
{\it Euclid} NISP will extend the wavelength range from to 900\,nm to 2\,$\mu$m, but  at a lower spatial resolution (compared to VIS) of $0.3\arcsec$ per pixel.

\emph{Euclid} and LSST  are thus strongly complementary for the key decision of what constitutes a valid celestial object (in the areas where the two surveys overlap). Two approaches for combining these data sets, with different levels of independence, are:

\begin{itemize}
\item Catalog-level combinations. For both surveys, objects would be detected independently and then matched.
The \emph{Euclid} VIS object catalog can be used to improve and calibrate the star-galaxy separation methods in LSST.
To identify so-called `ambiguous,' or unrecognized, blends \citep{Dawson2016}, for which the number of nearby overlapping objects cannot be determined in ground-based images, one can make
use of both \emph{Euclid} imagers:
VIS can spatially resolve peaks with separations of $\approx 0.3\arcsec$, and NISP can distinguish between objects with different NIR Spectral Energy Distributions (SEDs).

LSST object catalogs can provide confirmation of objects at the detection limit of \emph{Euclid} and identify faint and diffuse emission of objects whose full extent was not visible in the \emph{Euclid} images.
Furthermore, requiring corroborating detections across both surveys suppresses the contamination of spurious objects in the static catalogs, while providing additional time-domain information for transient and variable objects.

Beyond confirming objects, collaboration on catalogs can provide the opportunity to carry out ``forced photometry'' in both surveys, namely, measuring the flux in one survey based on the aperture of a source detected in the other survey. For resolved sources, a common aperture is not sufficient, and some form of modelling or PSF matching is necessary. For LSST, the VIS catalog could be used to measure the deblended flux in crowded fields based on the high-resolution \emph{Euclid} data.

\item Pixel-level combination. Object detection and characterization could be performed simultaneously on images from both surveys.
This is particularly important for separating blended sources, which will be about 40\% of all LSST objects \citep{Dawson2016}.
In contrast to catalog-level detection of blends, model-based or non-parametric deblending methods can benefit from a joint data set because of the extended wavelength range and resolution, enabling separation schemes that exploit color and morphological information.
Even for isolated objects, higher-level characterizations, such as photometry or shear measurements (\S\ref{sec:photoz} and \S\ref{sec:shearmeasurement}), can be performed at higher significance and fidelity when models are fit across several bands of both surveys, potentially employing priors on galaxy colors, sizes, and morphologies.

\end{itemize}

These two approaches are not mutually exclusive.
On the contrary, it appears prudent to perform the catalog-level combination first to establish whether the two surveys find consistent results for isolated objects.
Although this step will be limited to brighter objects, these are also the objects with the most precise shape and flux measurements.
Any inconsistencies (e.g. from image calibrations or the deconvolution from the PSF) should be seen as offsets in astrometric position, magnitude zeropoint, or ensemble lensing shear estimate from a matched galaxy sample.
These comparisons will establish whether common definitions, for detection thresholds, or photometric apertures, are being used, or how to convert measurements between the surveys. Such information will be vitally important for the wider community when they start using the public data products.

An expected source of inconsistency is due to unrecognized blends, which may show up as a single detection in LSST but may have several detections in \emph{Euclid} VIS (or NISP).  Furthermore, the NISP NIR centroids may not correspond to the LSST centroid derived from  the optical bands alone.
Passing along these measurements and their errors will enable downstream data processing and science algorithms to decide how to best characterize the blended group.

Once cross-survey inconsistencies are identified (if present), a joint
pixel-level analysis could extend the source catalogs in those domains
that are difficult for either one of the surveys but not for the
other. This would, of course, require the sharing of pixel level data
from \emph{Euclid's} VIS and NISP instruments and the multiband LSST
images (see, for instance, \S\ref{sec:dataproducts}).

Both approaches for joint object detection would be made considerably easier if data processing conventions have been coordinated from the beginning to share a common coordinate system and photometric standards (e.g. from \emph{Gaia}). While both \emph{Euclid }and LSST will likely use their own internal photometric system, there are obvious advantages to establishing a common set of photometric standards and making sure to exchange the standard photometric measurements between projects from an early stage. There are obvious cosmological gains from improved galaxy deblending, e.g., better shape and photometry measurements that result in higher surface densities of  galaxies useful for lensing, leading to improved statistical power and lowered systematics in dark energy studies. Beyond dark energy constraints, better deblending and the resulting higher surface density of lensed galaxies may also lead to a better understanding of the small-scale power spectrum, which allows for better constraints on the role of baryons, the nature of dark matter, and the neutrino masses.

\subsection{Photometric Redshift (Photo-z) Estimation}
\label{sec:photoz}

A variety of cosmological measurements (most notably weak lensing) from both projects will require accurate
knowledge of the redshift distribution of the source galaxies when split into
tomographic redshift bins. In the cases of LSST and \emph{Euclid}, the source galaxy samples will consist of billions of galaxies, and redshifts can only be obtained using photometric techniques. The resulting redshift estimates, generally referred to as photometric redshifts or photo-z’s, benefit from a large coverage in wavelength. 
The availability of a greater number of photometric passbands spanning a broader wavelength range will yield photo-z’s with higher precision; tighter redshift distributions will also be easier to calibrate.

LSST would benefit from the addition of the \emph{Euclid} NISP-photometric (NISP-P) passbands ($Y,J,H$) as they will extend wavelength coverage into the NIR, thereby breaking
degeneracies between redshift and the intrinsic SED of galaxies \citep{2010A&A...523A..31H,2000ApJ...536..571B}.
\emph{Euclid} depends crucially on optical multiband photometry from the ground. LSST will be superior to all alternatives in the southern hemisphere in terms of depth, number of photometric passbands, image quality, and area coverage.

The combination of data for photo-z's could be achieved at the catalog level. However, the photo-z's from both projects would be improved via a joint pixel-level analysis as described in \S\ref{sec:objectdetection}, as the effect of untreated blending on photo-z estimates is generally difficult to characterize. In addition, the photo-z probability distribution $p(z)$ for each object depends on the colors available and how they are measured (e.g. the relative weighting of the bulge and disk light in each galaxy).

A critical issue for both surveys will be calibrating photometric redshift algorithms, e.g., quantifying the error in the measured distribution of photo-z's (see \S\ref{sec:spectra_training}). Different techniques have been proposed to tackle this problem, including the use of a large, representative spectroscopic calibration sample \citep{2008MNRAS.390..118L,2015ApJ...813...53M} as well as
using angular cross-correlations between the samples of interest and galaxies with known redshifts   \citep{2008ApJ...684...88N,2015MNRAS.447.3500R}. All of these calibration techniques benefit from additional wavelength coverage of the photo-z and yield smaller errors when photometric redshift scatter is reduced \citep{2008ApJ...684...88N,2010ApJ...720.1351H}.

For LSST, the addition of the \emph{Euclid} NISP-P passbands would significantly reduce photo-z uncertainties at $z>1.5$. Likewise, LSST has the capability of covering two-thirds of the \emph{Euclid} survey footprint at greater depths than needed to achieve the required photometric redshift precision for \emph{Euclid} (see \S\ref{sec:overlap}). \emph{Euclid} operates with a fixed integration time in all bands, which will lead to non-uniform depths across its footprint due to the zodiacal light background (in the range of 0.6 magnitudes). While the \emph{Euclid} northern footprint will be covered by various telescopes at the depth needed to match the dark energy constraint driven requirement of the 30 source galaxies per square arcminute (VIS=24.5), LSST has the unique potential of producing photometry going significantly deeper than the  depth required by \emph{Euclid}.

Another advantage of using LSST, as opposed to a broader set of telescopes, is the relative uniformity of the PSF and properties of the data (including the data processing and calibration) across the passbands, facilitating a more homogeneous photometric calibration. Finally, LSST will perform many more visits per area than needed by \emph{Euclid}, allowing for the selection of the best subset for photometric uniformity (e.g., based on homogeneity of the PSF, delivered image quality, and atmospheric transparency). The additional $u$-band from LSST would also significantly reduce photo-z uncertainties at low redshift for \emph{Euclid}, which is useful for a proper treatment of a number of photo-z related systematics including galaxy intrinsic alignments  \citep[for a recent review of intrinsic alignments, see][]{2015SSRv..193....1J}.

\subsection{Shear Measurement}
\label{sec:shearmeasurement}

Cosmological weak lensing measurements require the ability to accurately infer the ensemble
statistics of the shear field, i.e. the weak lensing distortion as a function of position, and its
spatial correlation. Typically, the inference of the shear field is based on estimating the
weighted averages of individual galaxy shapes (for some definition of shape whose ensemble average has a well-defined
relationship to the shear).  The main challenge is to deconvolve the observed galaxy shapes from the influence of the PSF, which can vary temporally, spatially, and as a
function of wavelength.

For weak lensing measurements, three levels of coordination are advantageous. First,
the cross-correlation of the shear catalogs from the different surveys would help us expose and understand  any residual survey-specific systematics. Second, matching the object catalogs of both surveys, and then comparing the shear signals around a common set of foreground lenses (e.g. clusters) could reveal differences in the relative shear calibration. Finally, a
joint pixel-level measurement would be more robust against the problem of objects blending together (see, e.g., \S\ref{sec:objectdetection}), which is beneficial for LSST because of the wider PSF. Also, this could account for color variations within galaxies, which needs to be accounted for with \emph{Euclid} due to the single VIS filter for shape measurements \citep{2012MNRAS.421.1385V, 2013MNRAS.432.2385S}.

Methods to measure the ensemble weak gravitational
lensing shear distortions from PSF-corrected galaxy shapes typically need to be calibrated on simulated images containing a known shear \citep{henk2017}.
Although advances have been made to self-calibrate, using shear methods based directly on the survey data \citep{Huff2017,Sheldon2017}, additional test data need to be generated to independently verify the accuracy of the calibration \citep{henk2017}. Shear estimation methods that have been devised to avoid shear calibration biases that most shear estimation methods suffer by design \citep[e.g.,][]{bernstein16} require deep data to build their priors, and still require realistically complex image simulations to confirm that they are truly unbiased, similar to the needs of self-calibration methods.
Creating such realistic simulations of complex galaxy morphologies requires
images of the true sky with higher resolution and that are deeper than the survey images.
Calibrating the shear measurements of \emph{Euclid} could require an appreciable fraction of the \emph{HST} archive to statistically account for the impact of color gradients in galaxies \citep{2012MNRAS.421.1385V, 2013MNRAS.432.2385S}.

Calibrating shear measurements for LSST will
require a similar data set, like the \emph{Euclid} imaging or the \emph{Euclid} shear
measurements, especially from the \emph{Euclid} deep fields, which will be of comparable depth to the LSST lensing sample.
Calibrating multiplicative bias in shear measurements can
potentially be done via external measurements of the lensing of
the cosmic microwave background, or possibly LSST supernovae (but only at the level of a few percent; \citealt{2016arXiv160701761S}). This would be sufficient for some, but not all, of the tomographic redshift bins of \emph{Euclid} and LSST; the development of robust shape measurement methods will need to continue in order for both surveys to reach their shape measurement systematics requirements.

\subsection{Cluster Mass Estimates}
\label{sec:cluster_cosmology}

LSST and \emph{Euclid} will constrain cosmological parameters using galaxy clusters, primarily measuring the shape and evolution of the halo mass function.  To harness the power of this approach will require a precise and accurate cluster mass calibration \citep{Cosmic_Visions}.  Cluster weak lensing measurements are clearly the best opportunity to deliver the required mass estimates for ensembles of clusters \citep{henk2015,WtG4,Krause_Eifler}. The combination of \emph{Euclid} and LSST data will also help with cluster detection.

For accurate cluster masses from weak lensing, it is essential to separate the lensed background population from foreground  galaxies, especially cluster galaxies.  The combination of LSST and \emph{Euclid} provides more robust photometric redshift estimates, allowing one to  minimize the roles of external priors and calibration. From a \emph{Euclid} perspective, the combination with multiband, deep optical photometry is essential for cluster mass estimates at all redshifts in order to identify background galaxies.  From an LSST perspective, the addition of NIR photometry enables precise photo-z's beyond $z\sim1.4$, and thus an increased density of background sources with secure redshifts for clusters at $z\gtrsim 1$.

\begin{figure*}
\begin{center}
  \includegraphics[width=0.7\textwidth]{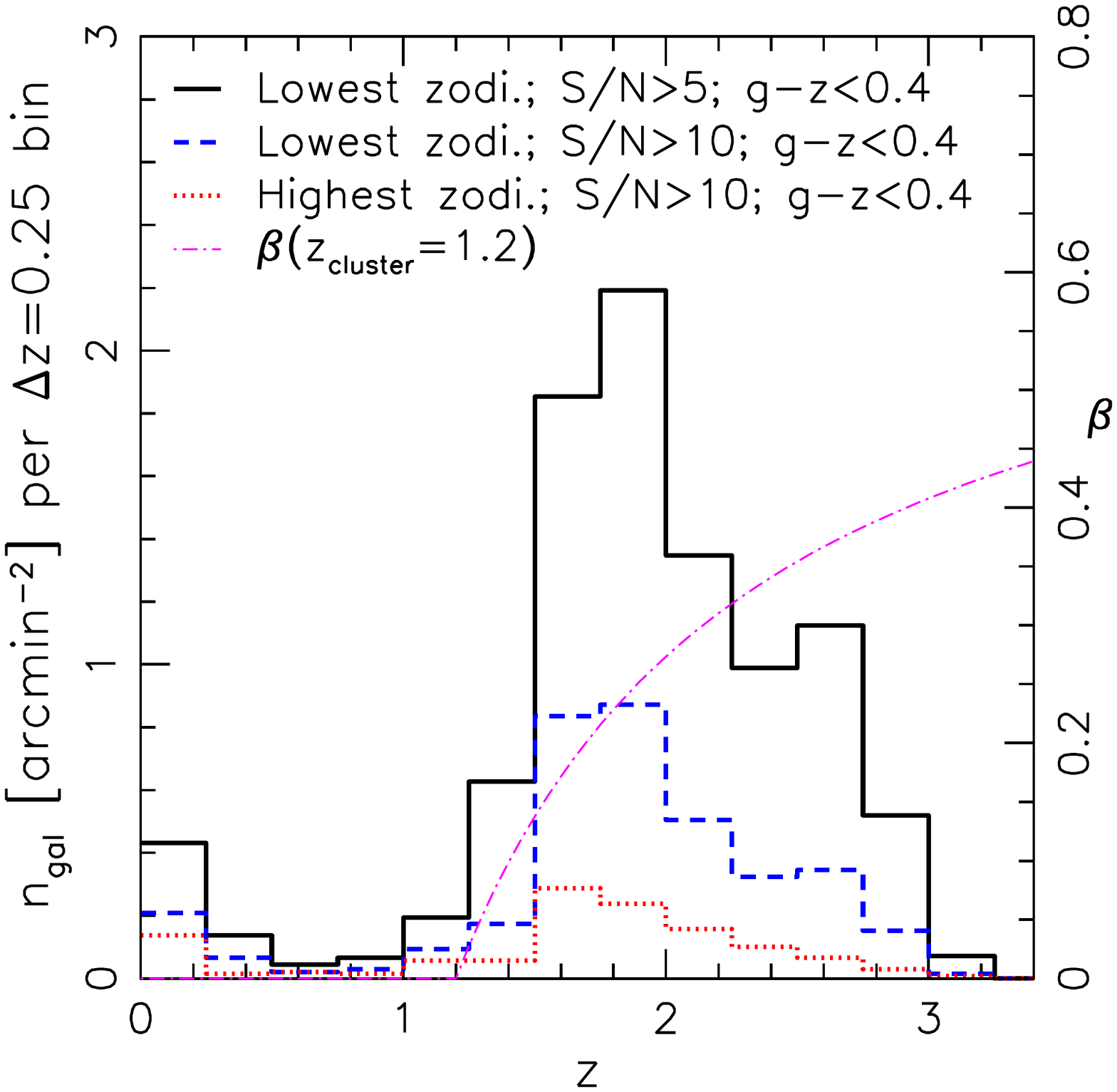}
\end{center}
 \caption{Weak lensing studies of clusters at
\mbox{$z_\mathrm{cluster}\simeq 1$--$1.3$}, require source galaxies at redshifts \mbox{$z\gtrsim 1.5$}, where the  lensing
efficiency \mbox{$\beta$} (shown in magenta for a
cluster at \mbox{$z_\mathrm{cluster}=1.2$})
is high.
Optical color selection from deep
LSST photometry (here \mbox{$g-z<0.4$}) is sufficient for selecting
most of these distant background galaxies  \citep[see also][]{schrabback16}.
The plotted histograms show the expected source density for \emph{Euclid} based on
CANDELS/COSMOS photometric redshifts \citep{skelton14}. While the source
density is low at low ecliptic latitudes with high zodiacal background (red dotted histogram, $|b|=15^\circ$),
it increases at high ecliptic  latitudes, where the Euclid data are deeper
(blue dashed histogram).
Compared to the default analysis using galaxies with a VIS signal-to-noise ratio
\mbox{$S/N>10$}, a further \mbox{$\times 2.6$} boost in the high-$z$
source density (solid black histogram) could be achieved with
advanced shape measurement techniques which yield robust results  for
galaxies with
 \mbox{$S/N>5$} \citep[e.g.][]{bernstein16}, in areas where deep LSST photometry is available for the color selection (see \S\ref{sec:coverage}).
\label{fig:zhisto}}
\end{figure*}

The combination of LSST high signal-to-noise ratio (S/N) optical photometry and \emph{Euclid}
space-based shape measurements is well suited to push weak-lensing studies
to higher cluster redshifts \mbox{$z_\mathrm{cluster}\simeq 1$--$1.3$}.
The required background source galaxies at \mbox{$z\gtrsim 1.5$} are still
well resolved in the Euclid VIS data for shape measurements.
Even if
 individual galaxy photo-z's may be too noisy,
the faint high-redshift tail of the expected galaxy distribution can still
 be selected from deep LSST colors  (and calibrated on the deep fields)  with low to moderate contamination from the foreground and
cluster galaxies
 \citep[see Figure \thinspace\ref{fig:zhisto} and][]{schrabback16}.
This strategy is particularly effective (i) at high ecliptic latitudes, where the
\emph{Euclid} data are deeper because of lower zodiacal background, (ii) where deep
photometry is available from LSST, and (iii) if new shape
measurement techniques, which yield robust results  for
galaxies with
lower S/N, are employed \citep[e.g.][]{bernstein16}.

The combination of LSST and \emph{Euclid} data will also play an important role in selecting clusters based on photometry, for example with the redMaPPer algorithm \citep{redmapper}, especially towards higher redshifts (also see Sect.~\ref{sec:cluster_astrophysics}). This will make the photometric data from \emph{Euclid} and LSST especially synergistic with other surveys across the electromagnetic spectrum, e.g., X-ray (eROSITA) and Sunyaev-Zel'dovich (SZ) surveys like AdvACT \citep{henderson16} and SPT-3G \citep{benson14}.

\subsection{Transients and Variable Sources}

The  transient survey of LSST is one key observational scientific difference with \emph{Euclid}. The   \emph{Euclid} wide survey will be covered in only a single visit, making it inadequate for transient searches (the \emph{Euclid} deep fields will have multiple visits over six years).  However, transient and variable source searches  could still be enhanced through coordination. For example, cosmological analyses based on Type Ia supernovae (SNe Ia) from LSST would benefit from NISP data on the host galaxies, e.g., \emph{Euclid} could provide spectroscopic redshifts, via NISP spectroscopy (NISP-S), for a fraction of LSST SNe Ia host galaxies, helping to remove possible systematic biases when using photometric redshifts alone \citep{2014AJ....147...75O}.

Beyond improving redshifts, the \emph{Euclid} data would help improve the classification and photometry of LSST transients. Firstly, \emph{Euclid} could provide serendipitous NISP spectroscopy for many active (long-lived) LSST transients, e.g., superluminous supernovae \citep{inserra17}. The low resolution of the \emph{Euclid} grisms are still sufficient to resolve the broad spectral features we expect from these objects and thus provide a definitive classification for a wide variety of serendipitous objects. Such a random (``unbiased'') sample provides an excellent training set for the LSST transient classifiers and is complementary to planned, dedicated spectroscopic surveys of LSST transients.
This will be particularly relevant in the first few years of overlap, when LSST will be spending most of its supernova-specific survey time building up a library of light curve templates with which to perform classifications later.
This will be timely in that \emph{Euclid} will have been operating for long enough that systematics should be under control and the
coordination of fields, and/or including a trigger for follow-up, will be of great use in improving classifications.

Moreover, for SNe Ia, even a few serendipitous  NIR data points overlapping with the LSST light curve (NISP-P and NISP-S) could improve the model fitting of these joint data, thus producing better distance estimates for a subsample of LSST SNe. That is, SNe Ia are better ``standard candles'' in the NIR \citep{2008ApJ...689..377W}. It is well-known that SNe Ia have a second light curve maximum in the NIR, thus extending their lifetime (in the observer frame) and increasing the opportunity for serendipitous detections by \emph{Euclid}.

Second, the high resolution \emph{Euclid} VIS imaging will provide important extra information about the host galaxies. Such information can help better inform the difference imaging required to detect transients, e.g. by convolving the higher resolution \emph{Euclid} image to match the lower resolution LSST template.
This would not be a simple task (due to the different passbands and PSFs, and the chromatic effects from the wide \emph{Euclid} PSF), so a coordinated effort between projects would be needed to unlock this science benefit.
Moreover, there is evidence that SNe Ia can be improved as ``standard candles'' by correcting their peak magnitudes by the local star-formation rate \citep{2015Sci...347.1459K,2015ApJ...802...20R} at the location of the SN within the host galaxy (e.g. by using local measurements of the galaxy colors enabled via the high resolution \emph{Euclid} imaging from VIS and NISP). Likewise, the extra information on the overall color and morphology of the host galaxy will be important for making additional corrections to the standardization of SNe Ia \citep[e.g.][]{2010ApJ...722..566L}.

Finally, there are interesting cosmological synergies between the shear weak lensing measurements from \emph{Euclid} and LSST, and the magnification weak lensing measurements possible from the LSST supernova sample \citep{2017MNRAS.465.2862S}. Such comparisons provide interesting systematic checks of the multiplicative bias, as well as possibly improving the signal from the combination of these measurements; for example, \citet{2017MNRAS.465.2862S} predict that the LSST SNe Ia-galaxy spatial cross-correlation function can be detected to high significance and can be used to provide additional constraints on cosmological parameters. It will be interesting to measure the spatial cross-correlation function of LSST SNe and \emph{Euclid} spectroscopic galaxies as any systematic uncertainties would be different between these two surveys.

The LSST supernova survey will allow for novel tests of cosmology, including tests of the isotropy and bulk flows. Using  bright tracers as a probe of the underlying structure and structure formation requires both an accurate estimate of the distance modulus (which will be improved with the NIR addition from \emph{Euclid}) and from more precise host galaxy redshifts, which \emph{Euclid} can provide. Work is underway to test the various cadence strategies of LSST on these novel probes, but coordination with \emph{Euclid} will enable this novel science with both telescopes.

The coordination discussed in this section applies to both the ``wide'' and ``deep'' fields planned for both experiments. For the wide surveys, this coordination may happen naturally, although care must be taken to ensure that the complementary survey strategies remain the default. However, more direct, active coordination may be required to ensure the LSST and \emph{Euclid} deep fields overlap both in area and time. Moreover, LSST and \emph{Euclid} could consider synchronization of  the observation of common deep fields, thus maximizing the number of unique, well-spaced, epochs. Such an active coordination would greatly benefit a range of time-domain studies, e.g., reverberation mapping of actoive galactic nuclei AGNs  \citep[AGNs; see][]{2009ApJ...705..199B}.



\section{Other Areas of Coordination}
\label{sec:other_benefits}
In this section, we explore the benefits of coordination that go beyond the cosmological goals outlined in the previous section.

\subsection{Spectroscopic Training Sets}
\label{sec:spectra_training}
\emph{Euclid} and LSST are exploring two routes toward spectroscopic calibration of their photometric redshifts.  One method is to calibrate the color-redshift relation using a spectroscopic sample of galaxies with an extremely high completeness; see \citet{2015ApJ...813...53M} who explored this strategy for \emph{Euclid}. This method is also explored in \citet{2017MNRAS.469.1186S} and \citet{2017MNRAS.469.1205S}.  The alternative is to measure the angular cross-correlations between the photometric samples and large, wide-area spectroscopic surveys. The true redshift distribution for the photometric sample is then reconstructed from this cross-correlation \citep{2015APh....63...81N,2015MNRAS.447.3500R}. The requirements on the spectroscopic samples needed by both LSST and \emph{Euclid} will be similar because of their overlap in area, redshift range, survey depth and calibration accuracy required.

For the color-redshift method, \emph{Euclid} and LSST have similar requirements on sample purity, quality, and knowledge of the selection function. At present, the  Complete Calibration of the Color-Redshift Relation \citep[C3R2;][]{2017arXiv170406665M} project aims to be complete to $RIZ < 25$ using spectroscopic data from Keck and VLT (located in several LSST deep fields including CDFS, SXDS, and COSMOS). This should be sufficient for the \emph{Euclid} wide survey, which will reach to $RIZ < 24.5$ AB magnitudes. Although not as deep as the LSST lensing sample of $i < 25.3$, this \emph{Euclid} training set will cover a significant fraction of the LSST color-color space for galaxies, so that a smaller number of additional spectra will be needed to calibrate LSST.

Based on previous studies, \citet{2015APh....63...81N} estimate that $\sim 20,000$ spectra in total would be needed to calibrate the LSST color-redshift relation down to the LSST weak lensing imaging depth. The  planned C3R2 sample is comparable in size, and the cosmic variance across six distinct 1 deg$^2$ fields (as envisioned for C3R2) would be comparable to that from the fifteen 0.1 deg$^2$ fields established as the LSST requirement in \citet{2015APh....63...81N}. Therefore, C3R2 should be sufficient for both surveys.

For calibration via cross-correlations, both LSST and \emph{Euclid} will have similar requirements on the total sky area, spectral density, and redshift range of the surveys.  As described in \citet{2015APh....63...81N}, LSST calibration using this technique will require at minimum 100,000 objects distributed over multiple widely separated fields of several hundred square degrees, spanning the full redshift range of the weak lensing samples used for dark energy analyses. \emph{Euclid} requirements have been estimated to be 0.4 galaxies deg$^{-2}$ per $\Delta z$=0.05 bin for $0 < z < 6$ over an area of at least 2000 square degrees, corresponding to 96,000 galaxies. Such samples should be available from overlap with the planned DESI and 4 m Multi-Object Spectroscopic Telescope (4MOST\footnote{{\tt https://www.4most.eu/}}) galaxy redshift surveys, where {\it Euclid} NISP-S can fill in the `redshift desert' (at $z=1.5-2.5$) which is difficult from the ground. Moreover, DESI and 4MOST should have ancillary programmes allowing the targeting of hundreds of thousands of targets across the overlap of these many surveys (e.g. the intersection of LSST, \emph{Euclid}, 4MOST and DESI).

More work is required to develop robust cross-correlation methods and demonstrate that they can in fact reach the calibration requirements of these dark energy experiments; the needs are similar for both \emph{Euclid} and LSST. Both methods require extensive validation, pre-processing, and metadata tracking for the spectroscopic samples.  For example, the process of validating the quality of spectra obtained will be a common problem for the two surveys. It would therefore be wise to proceed in partnership and include the planned wide-field spectroscopic surveys in this effort.  Cross-checks will occur naturally within each consortium, but overall coordination will reduce the total demand for external data and provide a means of quickly validating results.

\subsection{Numerical Simulations and Supercomputing}

LSST and \emph{Euclid} will each  require extensive numerical simulations for
a range of tasks  including forecasting, developing analysis techniques, and
  the eventual analysis of the observational data.  Carrying out
the simulations, transforming them into synthetic sky maps, validating
the results, and serving the data  in an easily accessible way are all
major efforts; they  require large computing and  storage resources in
addition  to  a  sufficient  workforce to  develop  the  modeling  and
analysis pipelines.

Quantifying  the  supercomputing   resources  required  for  numerical
simulations for each survey is  a challenging process. LSST has around
30 expert members that have  been active in defining cosmological
software  and hardware  infrastructure requirements.
Two reports have been delivered that define the computational requirements  for LSST cosmology for the years 2016-2022.
However, these reports specifically do not include numerical  simulations in their
estimates, so further effort is needed to define the needs in this area.

The future large-scale computing needs for DOE High Energy
Physics (HEP) have been articulated in a joint DOE Advanced Scientific
Computing  Research  (ASCR)  /  HEP  Exascale
Requirements Review Report \citep{2016arXiv160309303H}. The report contains an
analysis of HEP Cosmic Frontier simulation and data analysis computing
requirements for the years 2020 to 2025.   The annual  simulation
requirements for all HEP cosmology  experiments are expected to be 100 billion to one trillion (current) core hours by 2025
on ASCR High-Performance Computing  (HPC) systems, approximately a quarter
of  the projected  total HEP  requirement. Although  the three  main ASCR
facilities are  expected to support this  requirement, possible increases in requirements over a timescale of 5-10 years will
necessitate periodic re-evaluations.

The number, resolution, and contents of the numerical simulations that will eventually be needed for LSST and \emph{Euclid} remain an area of study \citep[e.g.,][]{2017arXiv170706529H, 2017arXiv170704488S, 2016MNRAS.456L.132S}.  It is possible that the simulations required for covariance matrix calculations could be extremely challenging, with up to a million independent realizations needed per covariance matrix using a brute force approach \citep{2013MNRAS.432.1928T}. Although there are active investigations underway to find more intelligent methods for producing covariance matrices \citep[e.g.,][]{2017arXiv170307786F}, this remains an open question for both  LSST and \emph{Euclid}. It is anticipated that even with orders of magnitude reduction in the number of simulations required for one covariance matrix, the total number needed for all covariance matrices will still demand significant computational resources.

Moreover, the surveys also require computationally expensive large, high-resolution N-body and hydrodynamic simulations for synthetic sky maps used for end-to-end testing of analysis pipelines and investigating systematic effects at sufficient fidelity for these Stage IV experiments, e.g. determining the impact of baryons on the weak lensing power spectra using state-of-the-art hydrodynamic simulations. Although the computational requirements for running these simulations and storing their outputs is commonly accounted for, the resources required to perform detailed analysis, and to store the resulting data products, are not as readily acknowledged.  Given the likely scale of the numerical simulations and analysis needed
for  the cosmological  science  of LSST and \emph{Euclid}, it would be sensible to coordinate
efforts  to  define  common numerical  simulations to  ensure  that  as much of these simulated data can be used by both surveys. This may require
slight modifications  to some  simulation requirements, but  a modest
overhead  that enables  sharing  would be  preferable  to each  survey
trying  to  acquire the  resources  to  run,  store, and  analyze  all
numerical simulations separately.

Such coordination would build on existing efforts already underway within the surveys and within various countries.
Initial efforts  were made toward  formulating a  plan for
what numerical  simulations are required  as a function of  time within each consortium. This work will  be valuable within \emph{Euclid} and LSST for planning purposes and should form  a template for coordination with between projects.

The United Kingdom Tier Zero (UKT0) group has brought together scientists from a wide range of areas in astronomy and particle physics (including gravitational waves).
The  primary goal of this  group is to
explore ways  to collaborate  on High-Performance Computing  and High
Throughput Computing.  In particular,  one experiment will utilize the
Large Hadron  Collider grid computing  in the context of  \emph{Euclid}, LSST,
and  SKA  studies to  determine  if  this  is an  effective  computing
solution for  these surveys.
Likewise, on the US side, there is  a  Tri-Agency (NASA,  NSF, DOE),  Tri-Project
(\emph{Euclid}, LSST,  \emph{WFIRST}) Group (TAG)  that has US  representatives from
each  of the  agencies  and  projects.  In  2015,  the TAG commissioned  a
US-based task force  with representation from each of  the projects to
write a report on the current state of cosmological simulations within
the  projects and  identify  benefits  of coordination.   The
report  submitted in March  2016   noted  that  coordination  is
certainly beneficial  in many areas,  but the significant  work effort
required to produce detailed plans would require  a separate task
force to be formed.
In early 2017, this led to the creation of a   Tri-Agency Cosmological Simulation (TACS) task
force,  formed at the request  of the US project  leads for \emph{Euclid},
LSST,  and \emph{WFIRST}.   The task  force  has representation  from US  and
European \emph{Euclid} members in addition  to representation from members of
LSST and \emph{WFIRST}.  TACS will  investigate the logistics of coordinating
hardware   between   the  agencies   to   enable   a  more   effective
supercomputing infrastructure from running  the simulations through to
their analysis, storage, and sharing with
a   wider  community.    The TACS  will  investigate   coordination
opportunities  for flagship  simulations, lower-resolution simulation
suites, and synthetic  sky generation.
Perhaps most importantly, TACS will investigate methods for reducing overall computing and storage requirements for simulations and whether simulation modeling efforts for a wide range of observational systematics are sufficient or need further work.

There  is broad  scope for  coordination between  \emph{Euclid} and  LSST for
numerical simulations  and many avenues  are currently being
explored.  The  significant resources required to  run, analyze, store,
and host  the numerical simulations for  any one survey are  the prime
reason to put effort  into coordinating between surveys as
much as possible.

\subsection{Astrophysics}
\label{sec:noncosmo}

The combination of LSST and \emph{Euclid} will transform all areas of astrophysics beyond the cosmological measurements discussed in previous sections. We do not seek to provide a comprehensive review of these synergies in general astrophysical studies but provide a few interesting examples to illustrate the range of impact envisaged.

\subsubsection{Solar System Science}
 In the context of planetary science,
  the strength of LSST is the high-cadence multi-epoch astrometry
  and photometry catalog that will contain hundreds of detections for each of
   millions of solar system objects (SSOs).
  This will dramatically improve our knowledge of SSO populations
  through discoveries and orbit determination \citep{2009-Book-LSST},
  phase curves for composition study
  \citep[e.g.,][]{2012-Icarus-219-Oszkiewicz}, and
  3D shape modeling \citep{2005-EMP-97-Durech,
    2015-AsteroidsIV-Durech}.

 Although the \emph{Euclid} wide survey step-and-stare survey mode suggests that
  only one epoch will be acquired for each target, its observing
  sequence is casually well-adapted for solar system needs
  \citep{2017-AA-Carry}; astrometric shifts within a single visit can indicate SSOs.
  The repeated visible and NIR photometry (over an hour) will
  provide the colors of SSOs, used for spectral
  classification and compositional interpretation
  \citep{2013-Icarus-226-DeMeo}. In particular, the NIR colors
  are key to resolving the present degeneracy between several spectral
  classes \citep{2009-Icarus-202-DeMeo} based on only visible wavelengths \citep[as produced by the SDSS
    or \emph{Gaia}, see][]{2002-AJ-124-Ivezic,
    2012-PSS-73-Delbo}.

 Nevertheless, orbital determination,  which is crucial to pinpoint
  compositions at specific locations in the solar system, will rely on
  data from other facilities as the hour-long observations by \emph{Euclid}
  will not suffice
  to constrain SSO orbits. This point was a major
  limitation of the SDSS census of SSOs, which detected over 400,000 moving objects but only half were linked with  sources with known orbits
  \citep{2001-AJ-122-Ivezic}. However, a posteriori identification
  is possible when the known population increases
  \citep{2014-AN-335-Solano}, e.g. from LSST.

   The synergy between LSST and \emph{Euclid} is obvious even at catalog level; spectral characterization from \emph{Euclid} can be
  linked with orbits and 3D shapes from LSST. To do so, an efficient
 \emph{Euclid}-centric ephemerides cone-search tool must be available
  \citep[e.g.,][]{2016-MNRAS-458-Berthier}. This will enable
  multiple investigations  including studies of
  dynamical families and surface aging effect \citep{2015-Icarus-257-Spoto},
  the source region of near-Earth asteroids \citep{2016-Icarus-268-Carry},
  highly inclined populations \citep{2017-AJ-153-Petit},
  and the large-scale distribution of material linked with
  solar system evolution \citep{2014-Nature-505-DeMeo}.

There are also synergies on the operation and cadence of LSST and \emph{Euclid}.
  By performing observations simultaneously, the distance to SSOs can
  be determined by triangulation, providing accurate orbits from only a
  few observations \citep{2011-CeMDA-109-Eggl}.
Trans-Neptunian Objects (TNOs) at 45au will have a parallax of a few arcseconds, tying down the distances to a high
certainty. This increased parallax will also  significantly reduce the orbital
calculation uncertainties of the hundreds of new near-Earth objects (NEOs) that LSST will detect per night (if coordinated observations are possible). In terms of the different passbands, doing multiwavelength imaging with LSST and \emph{Euclid} at the same time will give
very important physical information regarding transient events such as asteroid collisions and comet outbursts.
Owing to the observing geometry of \emph{Euclid}, which points close to a solar
  elongation of 90 degrees, and tight planned schedule, practical
  implementation of coordinated observations will require LSST to observe \emph{Euclid}
  fields right after dusk and before dawn.

\subsubsection{Internal Analyses of Galaxies}

The internal properties of galaxies at $0.2\lesssim z \lesssim 1$ are now being studied in detail, but the number of galaxies for which detailed data are available is small.  By combining LSST and
\emph{Euclid} we will revolutionize such internal studies.     For decades, we have explored galaxies as if they were single `points' and
even carried out analyses of galaxy evolution in terms of gross photometric quantities such as stellar and luminosity functions,
and the correlation of light and mass with other properties such as internal kinematics.   A few studies, mostly from \emph{HST} \citep[e.g.][]{1999MNRAS.303..641A,2007MNRAS.380..571L,2008ApJ...677..970W,2012MNRAS.424.1852L}, showed it was possible to determine the internal stellar population content of galaxies through a
pixel-level approach, i.e., treating each pixel as a distinct object and determining its stellar population parameters, such as its stellar
mass, light-weighted age, and metallicitiy.    In this way, one can utilize all potential information about galaxies which can reveal the
formation history of a single galaxy based on its spatial location.  This technique also allows the creation of  stellar mass images of
galaxies.

The combination of LSST and \emph{Euclid} data at the pixel level will create internal maps of  star formation histories, metallicities, dust content, etc. This will be achieved through SED fitting to the matched pixels across a range of wavelengths. This approach has been applied to only a few hundred galaxies, but such an approach should be applicable to 10 million galaxies up to $z\sim1$. This approach requires the imaging data from the two surveys to be aligned to high accuracy, and likely processed together, to avoid any offsets between the two and to ensure that the same physical regions are probed by both observatories. This will require matching of PSFs and pixel scales, as well as a careful control of the S/N and photometric limits, on a pixel basis. This approach can also be applied to fitting parametric models within galaxies, e.g., bulges and disk components.

\subsubsection{Morphological Classification of Galaxies and Machine Learning}

Galaxies display a wide range of morphologies, which encode information about their
potential energy and angular momentum, as well as their
cold gas content, the influence of cosmic environment on galactic
structure, and the history of mass accretion and mergers from
which the galaxies were built. As morphology is considered a
fundamental property of the galaxy population and provides a window
into the origin and fate of individual galaxies, surveys of galaxies
like LSST and \emph{Euclid} are significantly enhanced through the
morphological classification of each object they discover.

LSST and \emph{Euclid} will provide images for billions of galaxies. Although each galaxy will have many parametric measurements (luminosity, shape, size),
the sheer size of this database will be prohibitive for manually (visually) classifying these galaxies using the traditional Hubble sequence. Moreover, the Hubble sequence may not be relevant at high redshift, where many galaxies are irregular.

Fortunately, new machine learning methods
can  rapidly characterize of billions of images
through so-called `Deep Learning' models. After training, based on a subset of manually classified galaxies (possibly using GalaxyZoo; \citealt{2017MNRAS.464.4420S, 2015MNRAS.450.1441D}), we can use the deep learning models to
efficiently process enormous data sets with approximately the
same fidelity as manual classification. The higher-resolution \emph{Euclid} data, especially in the deep fields, will serve as an ideal training set for both \emph{Euclid} and LSST.

\begin{figure*}
\includegraphics[width=1.00\textwidth]{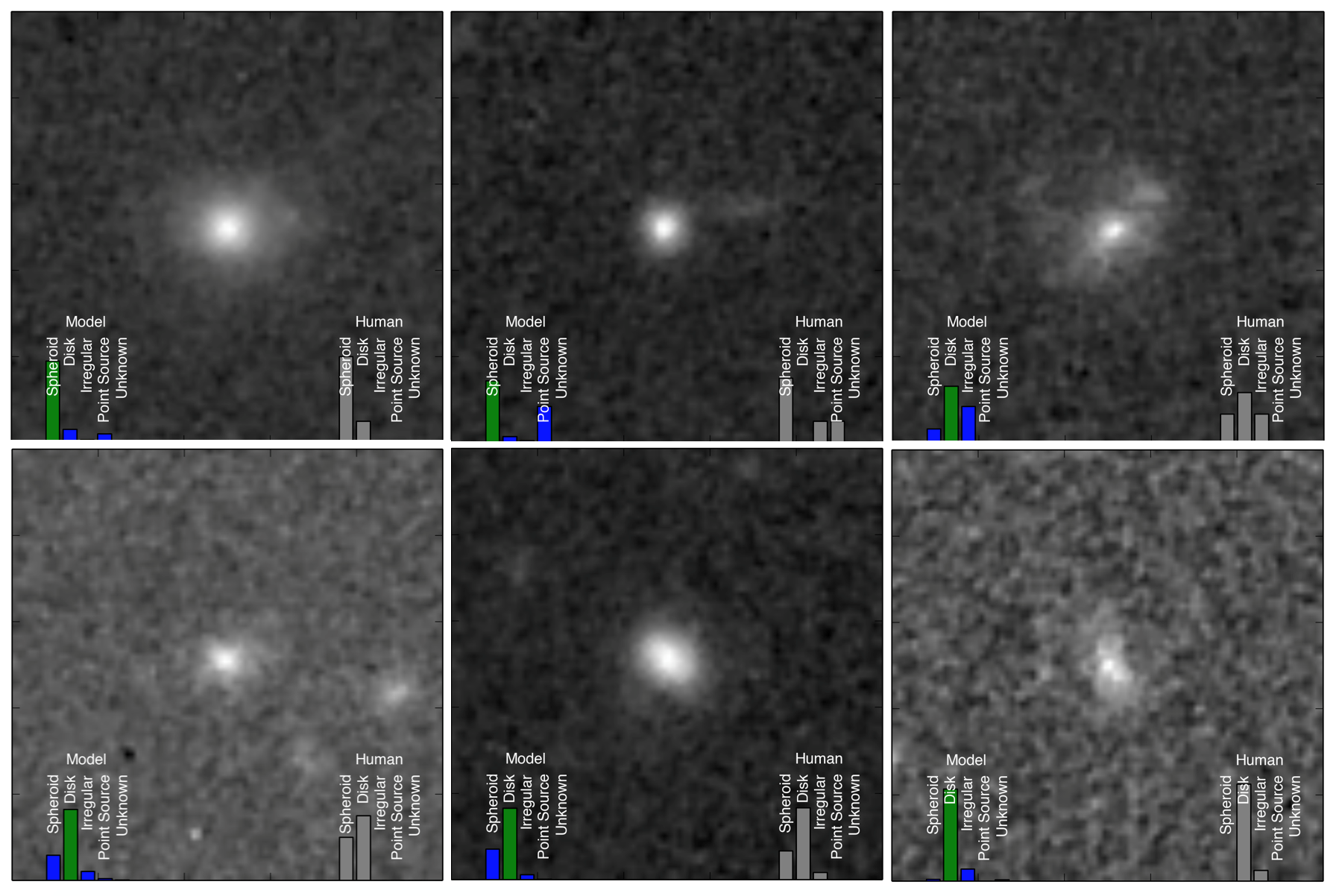}
\caption{Representative results of a deep learning model for galaxy morphological classification (Hausen \& Robertson 2018, in preparation) that will be applied to LSST and \emph{Euclid} imaging data. Shown are \emph{V}-band HST ACS images of galaxies in the CANDELS survey \citep{2011ApJS..197...35G,2011ApJS..197...36K}, with visual classifications provided by \citet{2015ApJS..221...11K}. The distribution of classifications assigned by human inspectors are shown as bar graphs in the lower-right corner of each image, reflecting the Spheroid, Disk, Irregular, Point Source, or Unknown classifications. The results of the deep learning model are shown as a bar graph in the lower-left of each panel, indicating the probability distribution of morphologies returned by the model. Green bars indicate when the most likely classification determined by the model matches the most frequent human classification.\label{fig:classification}}
\end{figure*}

Such deep learning models are already under development for
surveys like LSST and \emph{Euclid}. Figure \ref{fig:classification}
shows the results of a deep learning model (Hausen \& Robertson 2018, in preparation)
trained on the
\citet{2015ApJS..221...11K}
visual classifications of the CANDELS HST survey \citep{2011ApJS..197...35G,2011ApJS..197...36K}.
Shown are six
objects with inset comparisons of the Deep Learning morphological
classification probabilities (left corners) and those of the human-classified morphologies (right corners). As is evident from this
figure, Deep
Learning classifications can reproduce accurately the visual
classification schemes used by astronomers on an object-by-object basis,
but can be performed extremely rapidly and objectively.
Since Deep Learning is sensitive
to fine details in each image and uses this information to improve the
galaxy classification, a uniform processing of the data from both LSST and \emph{Euclid}
will reduce systematic uncertainties in machine learning galaxy classifications
that could otherwise arise from disparate treatments in noise, deblending, mosaicking, and artifacts. The combination of deeper LSST photometry and higher-resolution {\it Euclid} imaging could provide a new way of classifying galaxies, e.g. {\it Euclid}  focusing on the inner, brighter structures at high resolution and LSST detecting the outer, fainter parts including merger debris.

\subsubsection{Cluster Astrophysics and Dark Matter}
\label{sec:cluster_astrophysics}

In addition to their use as a cosmological probe (\S\ref{sec:cluster_cosmology}), galaxy clusters can be used to investigate the interplay of all major components of matter (dark matter, hot gas, and stars).  Weak lensing measurements with a high background source density can be used to map the total mass distribution in clusters, which has provided strong evidence for the existence of dark matter \citep{clowe06} and constraints on the dark matter self-interaction \citep{randall08,2015Sci...347.1462H}. Similar investigations can be performed with the joint \emph{Euclid} and LSST weak lensing analysis and would benefit from the increased source density.

As explained in  \S\ref{sec:cluster_cosmology}, the combination of {\it Euclid} and LSST data
 enhances cluster weak lensing studies at redshifts \mbox{$z\gtrsim 1$}, particularly at high ecliptic latitudes where there is reduced zodiacal background.
At faint magnitudes, the fraction of compact sources
increases, and while they are typically still resolved with \emph{Euclid}, they can be too small
for reliable shape measurements from the ground.
Hence, it is only through the combination of \emph{Euclid's} resolution and LSST's depth that the
additional statistical power of these faint and compact galaxies can be
incorporated into weak lensing studies, amplifying the total scientific
return.
In addition, the gain from a joint analysis including deep LSST photometry can be increased further
if advanced algorithms  that provide reliable
shape estimates from the \emph{Euclid} imaging and also for galaxies with a slightly lower
S/N than the default threshold S/N$>10$ (\emph{Euclid} will measure shapes at lower S/N) are implemented.
Promisingly,
\citet{bernstein16} already demonstrate accurate shape measurements for galaxies
with \mbox{$S/N>5$} for simulated data.

The extra depth in the weak lensing  source catalog
increases both the density and the average redshift of the source sample.
This boosts the statistical constraining power not only
for  cosmological weak lensing measurements, but
also other weak lensing studies, such as investigations of the mass
properties of galaxies as function of redshift \citep[e.g.][]{leauthaud12}
or the mass calibration of galaxy clusters  (see \S\ref{sec:cluster_cosmology}).
The largest relative gain in the source density occurs in the high-redshift tail
(see Fig.\thinspace\ref{fig:zhisto}).
These extra sources increase the
redshift lever arm over which `foreground' objects can be studied.

There are numerous motivations for studying galaxy clusters at $z>1$, including
their ability to tighten constraints on dark energy from clusters and
the longer redshift baseline for exploring the physics of the
intracluster medium and the evolution of galaxies. LSST data alone will be sufficient to identify systems with strong red sequences out to $z=1$; however, at $z>1$, the NIR data from \emph{Euclid} will be needed to detect and characterize such clusters. This is
underlined by \citet{2014ApJ...782L...3C}, who find that
secure (proto)cluster detection will require photometric redshifts with
at least $\Delta z/(1+z)\sim2.5\%$ precision at $z>1$.

\section{Details of Coordination}
\label{sec:tech_level}

The main data products needed for the weak lensing cosmological analyses of LSST and \emph{Euclid} are highly accurate photometric and shape measurements for as many galaxies as possible.
A number of algorithms have been developed (and are still being developed) to perform those measurements, some of which work on individual exposures, others on co-added images, ``co-adds'', that are aggregated from all suitable exposures for a given area of the sky.
The main trade-offs between those approaches are computational efficiency versus flexibility and accuracy.%
\footnote{In principle, it is possible to create an optimal, statistically sufficient ``likelihood co-add'' \citep{Kaiser2001, Zackay2015}, but the adaptation of this concept to survey data processing as well as downstream algorithms has not yet been successfully demonstrated.
}
For maximum accuracy, one may want to avoid the co-addition process as it can introduce artifacts from objects that are noticeably moving on the timescale of one to a few years; naive processing can also cause difficulties in interpreting the PSF across the boundaries where the number of co-added exposures varied or where some of the images have masked pixels. However, in practice, the data volume of LSST will probably demand that at least some data processing steps be performed on co-adds, e.g. object detection and initial characterization (\S\ref{sec:objectdetection}).

\subsection{Photometric redshift estimation}

The quality of photometric redshifts depends critically on accurate, high-S/N photometry in multiple bands.
In addition, galaxy photometry needs to be measured consistently across bands, i.e. a galaxy's stellar population needs to contribute to the integrated flux with the same weight in each filter. Even with a fixed measurement scheme, this color consistency is not guaranteed when the observing conditions (most importantly the PSF) vary between bands.

One way to account for that is by generating PSF-homogenized co-adds, where each contributing exposure has individually been convolved with an extra kernel to match a desired homogeneous output PSF.
If done for all exposures and all bands, integration apertures can be chosen in such a way as to optimize the S/N or to avoid contamination of adjacent objects.
The alternative approach is to determine apertures or fit models that account for the respective PSFs in each band, or even in each exposure,
as well as the presence of any adjacent objects (i.e., that handle deblending).

The following questions have to be answered for LSST and \emph{Euclid} independently. {\it i)} Which bands are used for source detection? {\it ii)} Can the PSF be characterized sufficiently well on co-adds or is that only possible on individual exposures? {\it iii)} Are fluxes measured on co-adds or on individual exposures? {\it iv)} Is the PSF variation accounted for by homogenizing the co-adds or by adapting apertures and models to the PSF differences between the bands?

The decisions made for these questions will in some cases affect further coordination efforts, all of which will at a minimum require the adoption of a common astrometric system.
Aperture fluxes are only consistent between both surveys if the PSFs have been matched to the same output PSF.
Model fluxes are consistent only if they adopt the same model parameterization and parameters for each object \emph{and} those models accurately describe the data. One must decide on a common aperture or model based on \emph{Euclid} VIS detections and subsequent object characterizations must then  be consistently applied to the LSST and \emph{Euclid} NISP images.
If, however, PSF-homogenized co-adds are adopted by LSST, a joint pixel-level analysis at the co-add level could negate any advantage of the higher spatial resolution of \emph{Euclid}. As a result, aperture and model fluxes of blended galaxies would remain susceptible to undetected leakage between the blends.

Once accurate multiband photometry is available, the estimation of photo-z's and their calibration is relatively independent of the project. Photo-z codes having a long history of development and have matured rapidly in recent years.
The requirements on scatter and outlier rates for \emph{Euclid} have been achieved in real-world datasets with existing codes  \citep{2016ApJS..224...24L,2010A&A...523A..31H}, although the demonstrations to date at \emph{Euclid}'s depth have utilized substantially more bands of photometry than \emph{Euclid}+LSST would provide.  These codes also meet requirements  for LSST on simulations \citep{2017arXiv170609507G}, albeit with the assumption of perfect knowledge of templates or very large representative training sets. Calibration requirements are also similar for both surveys (\S\ref{sec:spectra_training}).

Apart from algorithms, a key requirement for photometric redshifts for both projects will be enormous training and calibration samples that may only be obtained with highly multiplexed optical and infrared spectrographs on large telescopes.  Sharing and combining these calibration samples would certainly make sense but would likely not have a major impact on the data flow for either project.

 Previous works \citep[e.g.][]{2010ApJ...724..425D,2013ApJS..208....5N} suggest that using  high-resolution imaging data  (such as those \emph{Euclid} would provide in areas that overlap the LSST survey footprint) could reduce catastrophic redshift failures due to blending.  This is supported by as yet unpublished \emph{WFIRST} work (S. Hemmati \& P. Capak 2017, private communication), which suggests having a prior on galaxy shapes as a function of magnitude and other parameters would significantly improve LSST photometry.
The largest impact of the \emph{Euclid} NIR  photometry on LSST photometric redshifts will be for galaxies where $1.5<z<3.0$, for which the sharpest features in galaxy spectral energy distributions (SEDs), the rest frame $0.1216\mu$m (Lyman-$\alpha$) and the $0.4\mu$m (Balmer) break, are not constrained by the LSST filters.  This lack of spectral coverage decreases the precision of photometric redshifts and increases the outlier fraction.  We attempt here to provide some quantitative rigor to this claim based on simulated photometry, but leave a full analysis of the improvements of \emph{Euclid} and LSST photometric redshifts enabled by co-processing the data to a future work.

\begin{figure*}
    \includegraphics[width=1.00\textwidth]{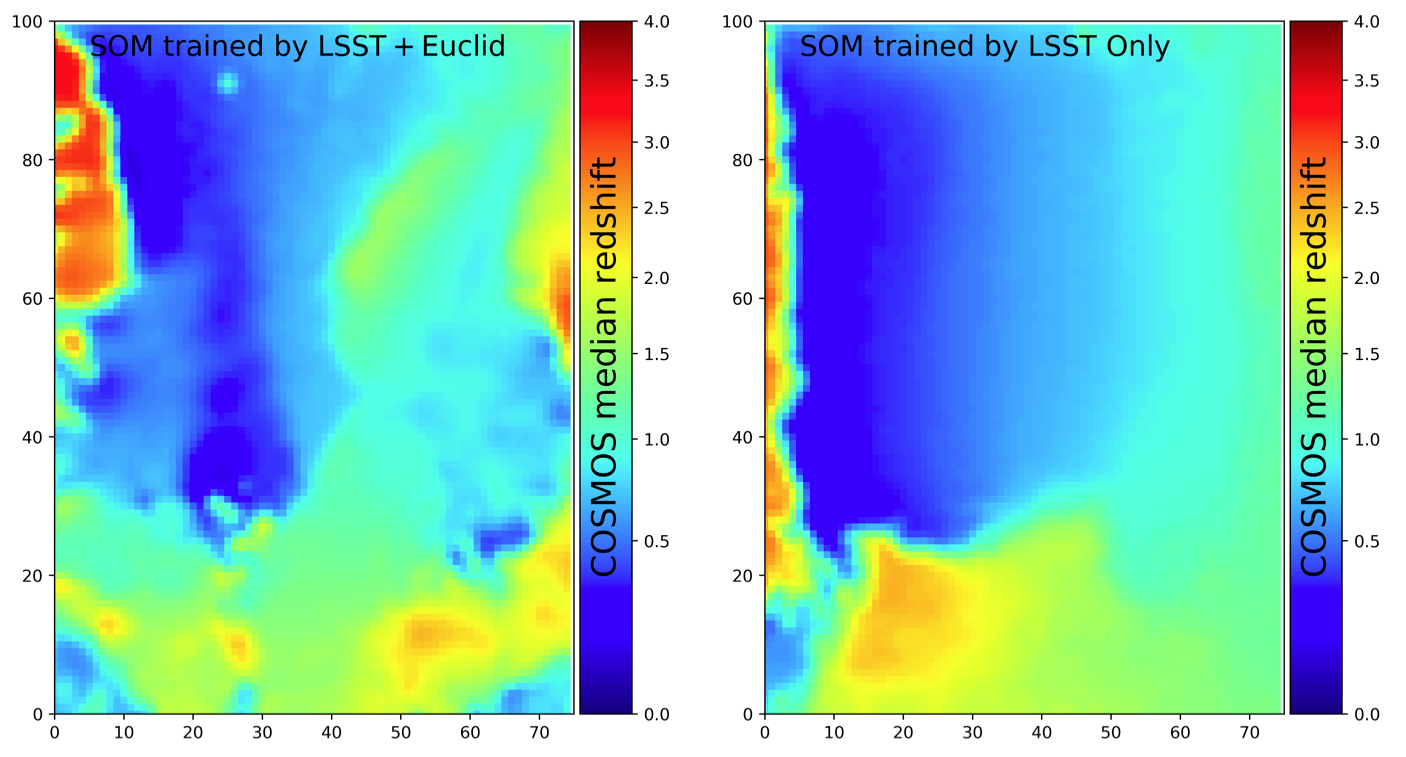}
     \caption{\small
    Self-Organizing Maps (SOMs) trained by colors of COSMOS galaxies (to the \emph{Euclid} depth) in the LSST+\emph{Euclid} filters (left) and the colors in the LSST filters only (right) color-coded by the median thirty-band redshift of the COSMOS galaxies mapped to each cell. SOMs are a class of unsupervised neural networks that reduce dimensions of data while preserving the topology. Each cell on these rectangular grids has a weight vector with the dimensions of the input data and therefore represents a point in the multidimensional color space (see \citealt{2015ApJ...813...53M} for details). The greater complexity of  features in the SOM shown on the left reflects the increase  in the information content when \emph{Euclid} bands are incorporated in the training. These features show that the photo-z resolution in a multidimensional color space has improved.
     }
     \label{fig:SOM}
\end{figure*}

\begin{figure*}
    \includegraphics[width=1.00\textwidth]{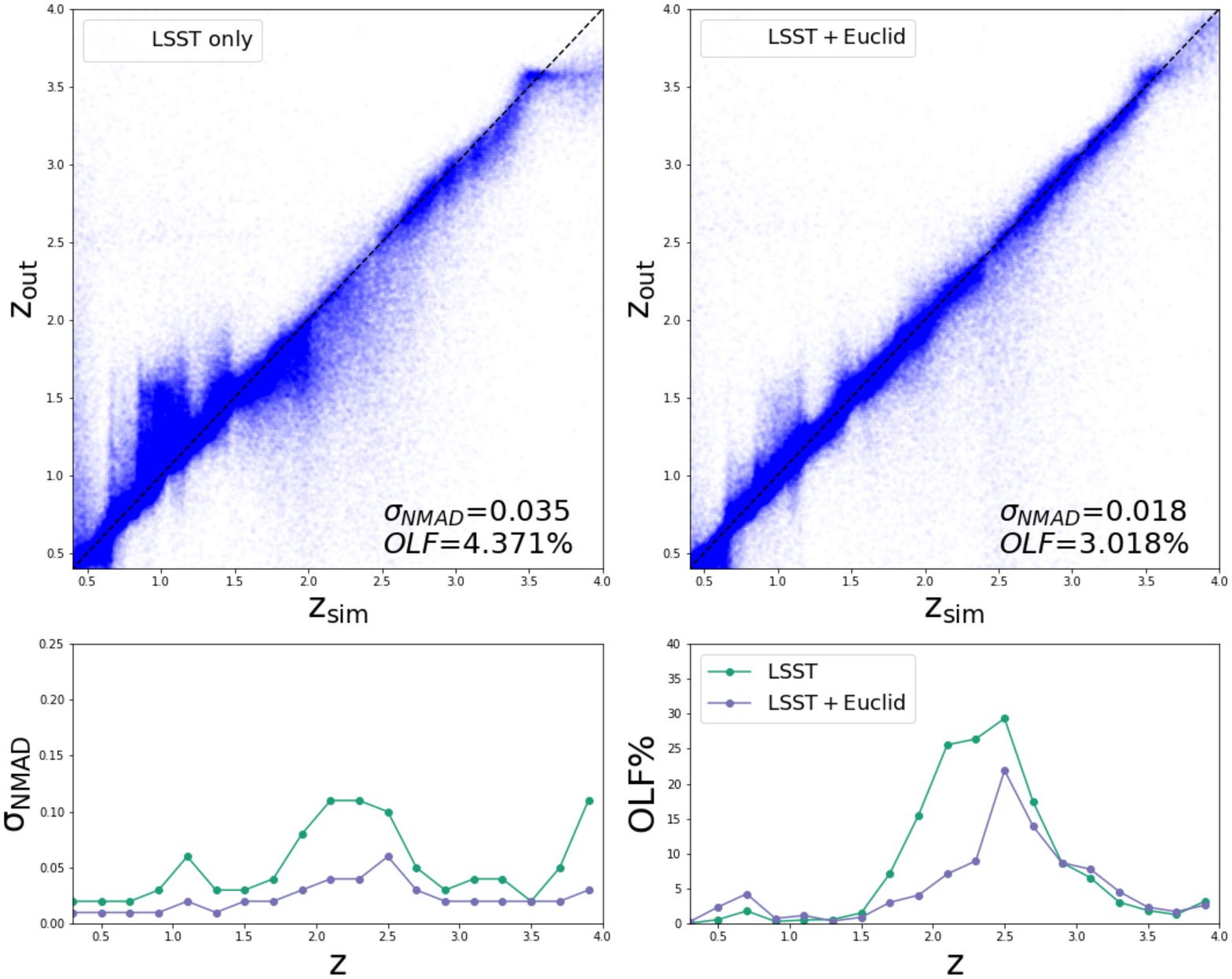}
     \caption{\small
   Top:  A comparison of simulated photo-z vs spec-z performance for LSST and LSST+\emph{Euclid} using the methodology described in \citet{2016arXiv160606374S}.  Clear improvement in the performance can be seen.  Bottom:  The $\sigma_{NMAD}$ defined as $1.48\times \rm median(|\Delta z|/(1+z_{spec}))$ and outlier fraction (defined as the fraction of objects with $|\Delta z|/(1+z_{spec})<0.15$) in redshift bins of 0.2 are shown for the simulation.  In these simulations, both the dispersion and outlier fraction improve by a factor of $\sim2$ between $1.5<z<3$.
     }
     \label{fig:photzspecz}
\end{figure*}

	First, we use the COSMOS data \citep{2016ApJS..224...24L} scaled to approximate the LSST+\emph{Euclid }data.  In Figure~\ref{fig:SOM}, we show a self-organizing map \citep{2015ApJ...813...53M} of the LSST and LSST+\emph{Euclid} data, which provides a 2D projection of the higher-dimensional color space.  One can clearly see additional features in the LSST+\emph{Euclid} color space compared with the LSST color space.  This indicates a significant increase in the information content.  Next, we simulate the photo-z performance using the methodology described in \citet{2016arXiv160606374S} but only simulate the end-result LSST and \emph{Euclid} photometry with estimated Gaussian noise and do not perform end-to-end image simulations.  In brief, the simulations start with the COSMOS2015 catalog and photometric redshifts \citep{2016ApJS..224...24L}.  We then fit a combination of \citet{2014ApJS..212...18B} galaxy and \citet{2011ApJ...742...61S} AGN templates modified with additional emission line ratios and dust obscuration to the non-stellar photometry.  This normalized template set is then used to estimate LSST and \emph{Euclid} photometry, which is then degraded to the expected sensitivity levels in those surveys.  In Figure~\ref{fig:photzspecz} we show a simulation of the expected improvement in the LSST photo-z performance for the LSST ``gold'' sample $i_{AB}<25.3$.  In these simulations, the scatter decreases by a factor of  $\sim2$ in the $1.5<z<3$ redshift range (the so-called `redshift desert') and $\sim30\%$ at other redshifts. The improvement of photo-z’s in the redshift desert provided by the combination of \emph{Euclid} and LSST will also increase the precision from photometric redshift calibration techniques that rely on spectroscopic cross-correlations/clustering redshifts, as tighter redshift distributions yield both smaller calibration errors and a decreased sensitivity to bias evolution \citep{2008ApJ...684...88N}.  This will correspondingly reduce systematic uncertainties in  the large-scale structure and weak lensing probes of cosmology from each experiment.
 However, the true performance is sensitive to the photo-z method, depth, data quality, selection function, and other properties of the LSST photometry and photo-z pipeline and so our projected improvement should be taken as a potential scaling rather than a definitive result (for instance, gains are smaller although still significant in simulations where magnitude priors are incorporated; S. Schmidt 2017, private communication).

  Although this improved performance will not directly apply outside of the overlapping area, it will provide statistically robust priors for the full survey.  For instance, the effects of blending on the photo-z distribution can be quantified as a function of LSST color and magnitude space and applied to the full survey.  Similarly, the improved photo-z distributions in the overlapping area will provide redshift priors for the full LSST sample where the infrared data are not available.

\subsection{Shear measurement}

As already discussed in \S\ref{sec:objectdetection}, joint analysis of LSST and \emph{Euclid}  images could be important to improve deblending issues in LSST, which has been identified as an important factor for LSST weak lensing  measurements, given the depth of the images. More generally, shear estimation hinges on highly robust and well-understood algorithms to estimate
the PSF and to remove the impact of image defects and  of the PSF on the galaxy shapes
to enable a robust ensemble shear estimate with systematics removed or properly marginalized over.
Here, the complementarity of LSST and \emph{Euclid} is primarily in the fact that the issues in PSF
estimation in the two surveys are quite different (and independent), image defects do not occur at the same levels, and
they experience different levels of chromatic effects (which cause the effective PSF to differ for
each galaxy or even for different points in the same galaxy if it has color gradients). For this reason,
many of their shape systematic errors can be treated as independent. \emph{Euclid} and LSST
will also measure shapes using different spatial weights, which, in addition to being a possible cross-check on systematic errors,
can be relevant in handling intrinsic alignment effects or in using intrinsic alignments as a probe of cosmology \citep{2016PhRvD..94l3507C}. For example, the combination of two shape measurements from the same sample of aligned galaxies can aid the mitigation of cosmic variance (similarly to \citealt{2009PhRvL.102b1302S}) when constraining  anisotropic models of primordial non-Gaussianity.

Because of this complementarity, one useful high-level coordinated analysis would be the cross-correlation of  shear fields in some region of the sky (see \citealt{2006JCAP...02..001J} for an early example of this type of analysis in the context of separate exposures within a survey). This
does not require coordination at the image or catalog levels.  Nearly all sources of coherent
additive biases should cancel out with such a cross-correlation, meaning that this joint analysis
should provide a cleaner shear power spectrum requiring less marginalization over nuisance systematics than auto-correlation.

To test for multiplicative offsets in the shear calibration of the two surveys would require the sharing of catalog-level products (e.g. per-object weights, photometric redshifts, and galaxy shapes).  For this test, we would select a sample of massive
foreground objects, such as photometrically detected galaxy clusters, and compare the galaxy-galaxy
or cluster-galaxy lensing amplitudes.  If each survey calculates the shear or surface mass
density using internal data products, a comparison of the results will not cleanly tell whether
there are disagreements in shear calibration or photometric redshifts.  However, if one survey uses
a matched sample of objects to infer the surface mass density around these lenses, and the other
uses the cross-matched catalog to adopt the weighting scheme and photometric redshifts from the
other survey, while using internal galaxy shape estimates, then a comparison of the inferred survey
mass density should agree in the absence of relative shear calibration biases. \citet{2017arXiv170704105A} offer a recent example of this type of comparison using  KiDS $i$ and $r$-band data with very different depths. All such
comparisons should be done at the level of inferred shears (or
inferred surface mass densities), not per-object galaxy shapes.  Per-galaxy shapes measured using
different algorithms should not necessarily agree depending on differences in weighting schemes and
resolution of the imaging data, so comparison must be done using the quantity that is really of
interest - the ensemble shear estimate.

\subsection{Weak Lensing S/N and Photo-z Accuracy: An Example}
\label{sec:example}

We provide here an example calculation (with some simplifying assumptions), demonstrating that the weak lensing S/N is increased due to the improved photo-z accuracy in the overlap area between \emph{Euclid} and LSST. Future efforts will do more complete calculations and full joint dark energy forecasts for the \emph{Euclid} and LSST combination. We include this calculation as a preliminary  demonstration of the power of combining these two surveys.

To this end, we consider the S/N for measuring the shear power spectrum from the \emph{Euclid} and LSST data and the cross-correlation spectra between sources in the common area. For this analysis, we first assume that one is not sharing data between the two surveys, but only use the final catalogs so that photo-z accuracy is the standard for each survey taken on its own.   The data vector is then

\begin{displaymath}
{\bf D}^{T}(\ell) = \{{\cal{C}}_{ij}^{EE}(\ell), {\cal{C}}_{ij}^{EL}(\ell), {\cal{C}}_{ij}^{LL}(\ell) \}
\end{displaymath}
where ${\cal{C}}_{ij}^{XY}(\ell)$ is the shear power spectra for the redshift bin combination $(i, j)$ at the multipole $\ell$ for sources in the the $X$ and $Y$ catalogs (with $X = E, L$ for \emph{Euclid} and LSST). The covariance matrix ${\bf \Gamma}_{ij}(\ell)$ can be computed as detailed in \citet{2016MNRAS.463.3674H}, to which we refer the reader  for further details. For a given $\ell$, we the define the S/N ratio as
 ${\cal{S}}_{+} = \left \{ \tilde{{\bf D}}^{T}      {\bf \Gamma}^{-1} \tilde{{\bf D}} \right \}^{1/2}$
where quantities with a tilde are computed for a given cosmological model and using the photo-z specifications for each single survey.

We now recompute the S/N under the assumption that the two surveys share photometric data so that the photo-z accuracy is better for the sources in common. The \emph{Euclid} and LSST shear catalogs will then be split in two parts depending on whether sources are in the overlap area. Both the theoretically expected values and the covariance matrix must be recomputed accordingly. We can then define a new S/N ratio as
${\cal{S}}_{\times} =\left \{  \hat{{\bf D}}^{T} \hat{{\bf \Gamma}}^{-1} \hat{{\bf D}}_{ij}   \right \}^{1/2}$
where now the hat quantities are computed with  improved photo-z accuracy parameters assuming a fraction $f_{E}$ of \emph{Euclid} sources are also detected and useful for lensing\footnote{We do not expect $f_{E} = 1$ since LSST has a lower expected $n_{eff}$ (number density of galaxies useful for weak lensing)  than \emph{Euclid}.} by LSST.

\begin{figure*}
    \includegraphics[width=1.00\textwidth]{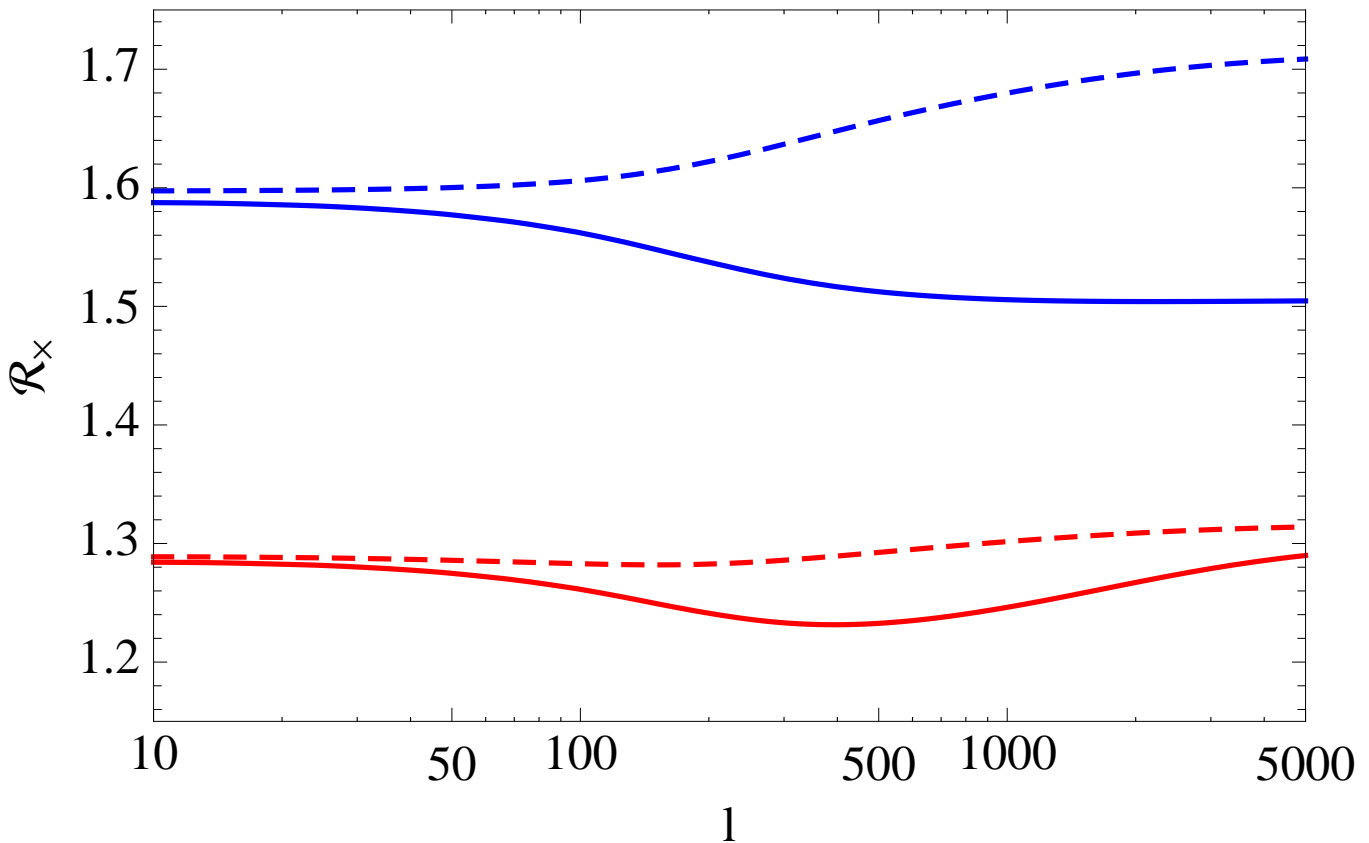}
     \caption{\small
S/N ratio of the measurement of the lensing data vector ${\bf D}$ defined in the text (including contributions from both the auto and cross-correlation power spectra) with and without photo-z improvement from survey combination.
Blue (red) lines refer to the case with $11000   (7000)$ square degrees of  overlap between the two surveys, while dashed and solid lines are for $f_{E} = 0.75$ and $1.0$, respectively. We use 100 logarithmically equispaced bins in $\ell$ over the range $10 \le \ell \le 5000$, and 10 approximately equipopulated redshift bins over the range $0 \le z \le 2.5$. We set $i = j = 6$ so that the results refer to the bin combination corresponding to the median of the galaxy samples from each survey. Changing the values of $(i, j)$ gives qualitatively similar results. }
     \label{fig:cardone2}
\end{figure*}

Figure\,\ref{fig:cardone2} plots the ratio ${\cal{R}}_{\times} = {\cal{S}}_{\times}/{\cal{S}}_{+}$ as a function of $\ell$ for the autopower spectra of the most populated redshift bins. We consider two different cases for the overlap area (namely, $7000$ and $11000$ square degrees) and for the fraction of \emph{Euclid} galaxies in common with LSST (i.e., $f_{E} = 0.75$ and $f_{E} = 1$). As it is apparent, the boost in the S/N from sharing data between the two surveys can be quite large even under the most conservative assumptions. Although we are well aware that a boost of a factor $b$ in the S/N does not automatically translate into a similar improvement in the constraints on the cosmological parameters, this preliminary result (obtained by ignoring the systematics and the intrinsic alignment contribution) clearly suggests that the improved accuracy in photo-z's coming from the joint use of \emph{Euclid} and LSST photometric data offers the fascinating possibility to significantly increase the amount of information from the two surveys.

\subsection{Survey Characteristics}

\label{sec:overlap}

\subsubsection{Sky Coverage}
\label{sec:coverage}
\begin{figure*}
    \includegraphics[angle=90,width=1.00\textwidth, height=1.00\textheight]{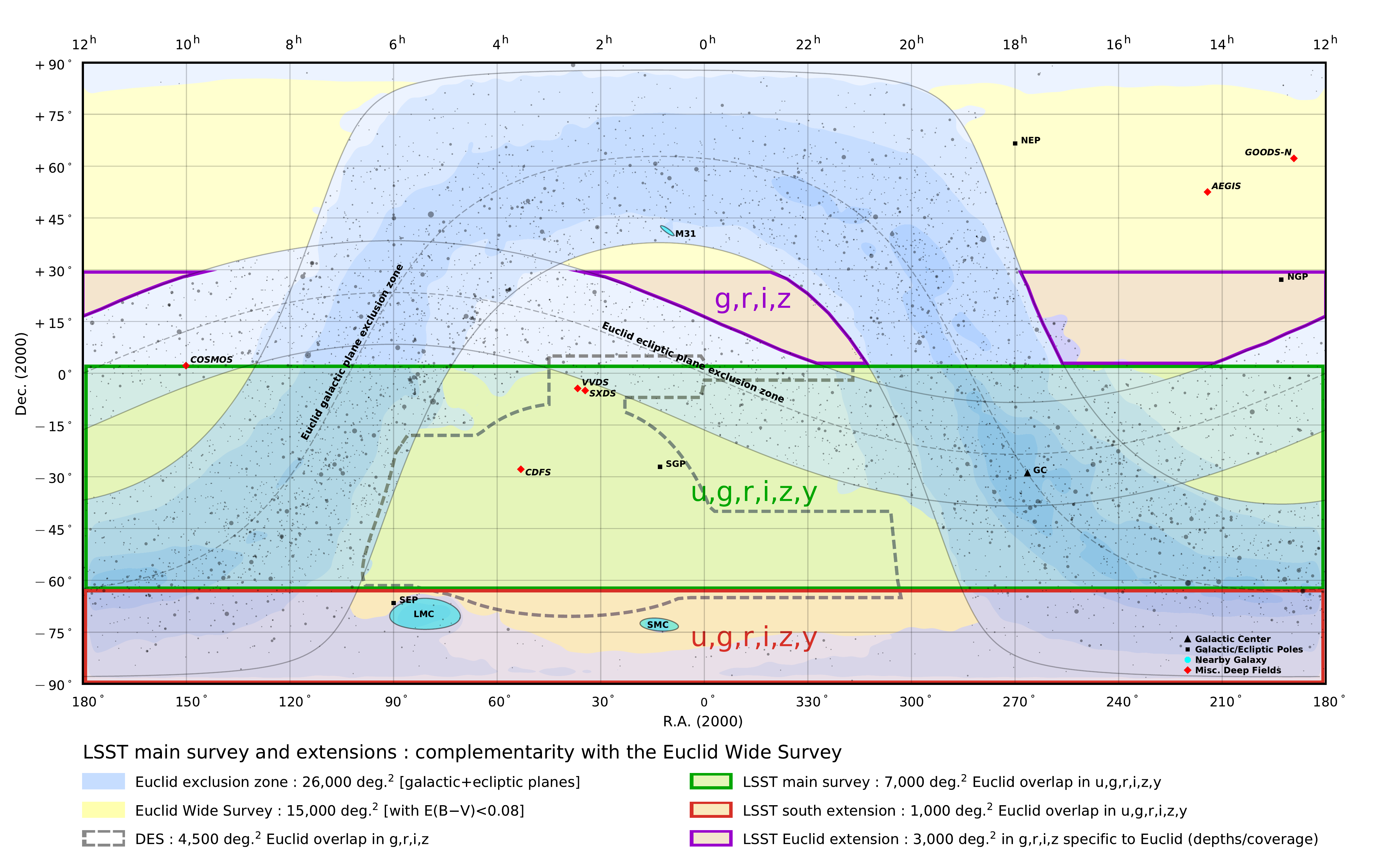}
     \caption{\small
     LSST survey areas and photometric bands and the \emph{Euclid} Wide Survey with its exclusion
     zone (blue: galactic plane + ecliptic plane + reddening). We indicate in the legend  the number of square degrees
     from the LSST surveys that overlap the \emph{Euclid} Wide Survey in the relevant photometric bands.
     }
     \label{fig:survey}
\end{figure*}

\begin{figure*}
     \includegraphics[width=1.00\textwidth]{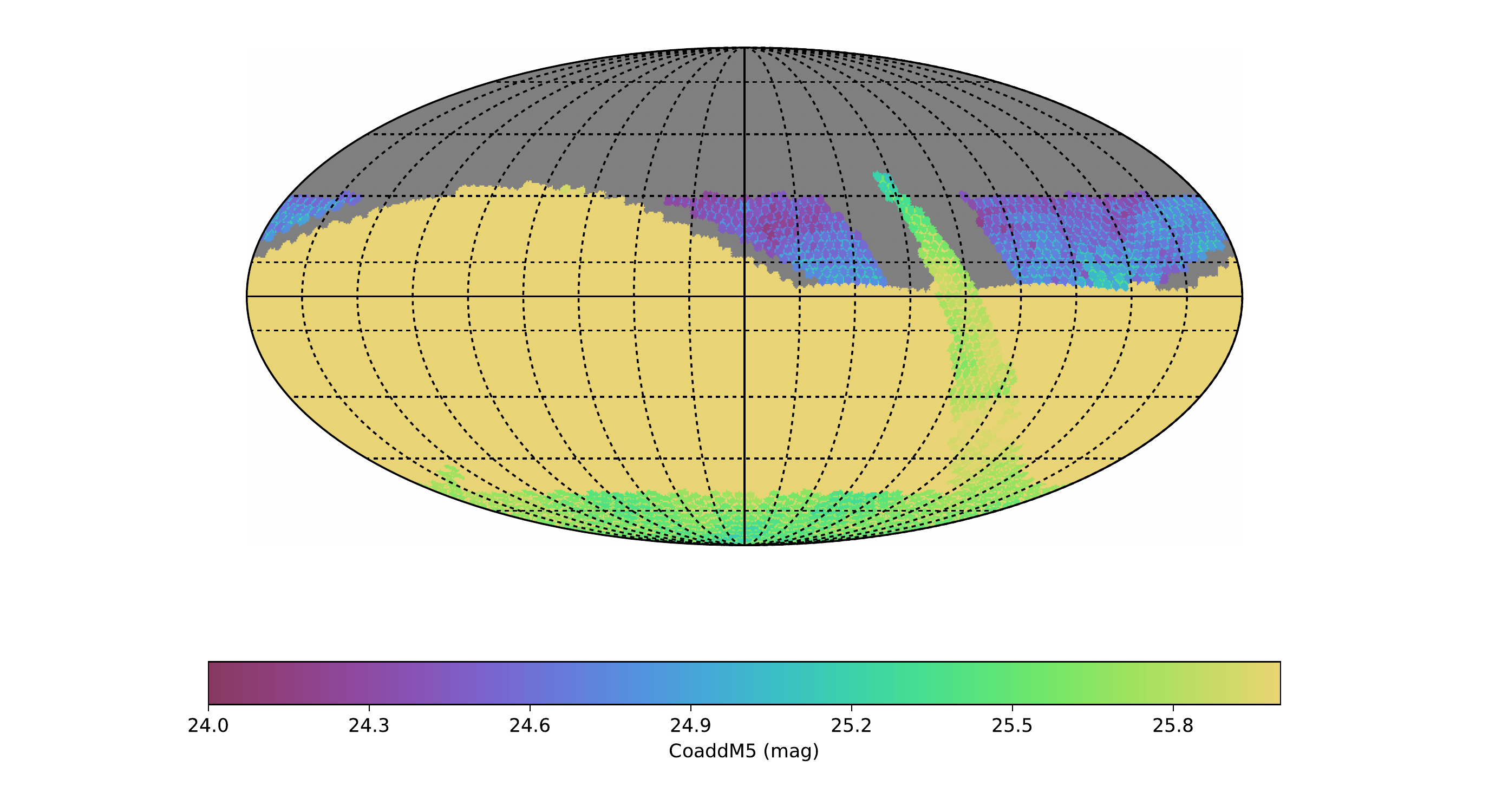}
     \caption{\small
       Simulation of an extended LSST survey at high latitude (declination above +2 deg. and up to +30 deg.)
     that increases the overlap between \emph{Euclid} and LSST by approximately 3000 square degrees (using the target  LSST depths described in the text). The decrease in image quality and the increase in atmospheric extinction for these high airmass observations reduces the effective exposure times at the high declination limit of this survey. A more detailed study will be required to determine how to minimize the impact of such effects on the LSST-\emph{Euclid} science cases, e.g.\  photometric redshift estimators. The simulation also shows the planned LSST ecliptic plane survey at northern latitudes for reference.
     }
     \label{fig:extend}
\end{figure*}

The main LSST survey will cover a total area of 18,000 square degrees located between an equatorial declination of $+2$ degrees
and $-62$ degrees to ensure all data contributing to its weak lensing science are acquired under an airmass of 1.4. This defines
a large sky area denoted by the green box on Figure~\ref{fig:survey}. There will be secondary LSST surveys such as the South Celestial Pole survey that will bring a shallower exploration of the
southernmost part of the sky (declination $-62$ to $-90$ degrees, the red box in Figure~\ref{fig:survey}) as well as the ecliptic
plane at the  northern declinations, up to and above +30 degrees (Figure~\ref{fig:extend}). LSST will use the same observing strategy for all of its
surveys with a pair of short fixed integrations (two 15 second exposures) per single visit of each sky area in all of its six available
photometric broad bands ($u,g,r,i,z,y$).

There are three key constraints to the visible area for \emph{Euclid}. First, stellar contamination, coupled to stray-light
constraints from the integrated light from the galactic plane, stops \emph{Euclid} from observing at low galactic latitude.
Second, with its constant integration time, \emph{Euclid} has to avoid areas with significant attenuation from galactic
reddening. Finally, the zodiacal light from L2 causes a significant background, which excludes the ecliptic plane (the S/N would
be too low for the fixed integration time, especially for the NISP). Therefore, to ensure \emph{Euclid} can achieve its
requirement of at least 15,000 deg$^2$ of extragalactic sky over the entire sky, the survey area is defined simply as
galactic latitude $>\pm25$ degrees, ecliptic latitude $>\pm15$ degrees, and reddening\,$E(B-V)<0.08$ (and up to
$E(B-V)<0.15$ to avoid holes or islands in, or near, the main footprint).

These different areas are shown in Figure~\ref{fig:survey}
on an equatorial coordinate projection, with the \emph{Euclid} Wide Survey shown in yellow, alongside the DES
footprint and other various points of interests. This simple definition presents the clear advantage of focusing on the best
parts of the extragalactic sky. There are, however, areas at lower galactic latitudes near $l = 180$ that \emph{Euclid} could
explore since requirements on scattered light might still be met; this is currently being explored by \emph{Euclid} using various
weight maps that draw a longitude-dependent exclusion zone.

LSST could reach further north than its present survey region due to its location ($-30$\,degrees latitude). Therefore, if we
wished to maximize the LSST-\emph{Euclid} overlap, nearly 11,000 degs$^2$ of the \emph{Euclid} wide survey could be covered by LSST (considering a longitude-dependent exclusion zone, not the depiction shown in Figure~\ref{fig:survey}),
or nearly three-quarters of the required 15,000 deg$^2$. This would require a new, currently unplanned, dedicated LSST
extension survey of 3000 degs$^2$ from a declination of +2 degrees to +30 degrees. Such an extension would however only
need to reach the modest depths required by \emph{Euclid} shown in Table~\ref{tab:depth}, and would exclude the $u$-band which would
otherwise require excessive integration times at such high airmass.

In Figure~\ref{fig:extend} we illustrate this extension with a simulation using the Operations Simulator
\citep{2014SPIE.9150E..15D} where we extend the LSST footprint survey with declination $<$ +30 degrees, |galactic
latitude| $>\pm$25 degrees, and |ecliptic latitudes| $>\pm$15 degrees. For this case, we allocate 43 visits spread
across the filters, excluding the $u$ band. Figure~\ref{fig:extend}  shows
the effective $r$-band depth achieved by this new survey. This extension only requires $\sim$222 hours (or 27 full
nights) to cover the additional 3000 degs$^2$ in the $g,r,i,z$ bands.

\subsubsection{Depth Requirements}
\label{sec:depths}
LSST will observe its survey areas in all of its six available photometric filters ($u,g,r,i,z,y$) using
a fixed integration approach (two 15 second exposures per single visit of each sky area), leading to different
depths across the various bands. There will be approximately 825 visits per field  area over the 10 year long
survey, leading to a net gain of more than 3 magnitudes over the single visit depths listed in Table~\ref{tab:depth},
assuming a point-source $5\sigma$ PSF photometry metric/performance \citep{2008arXiv0805.2366I}.

The \emph{Euclid} Wide Survey average depth requirements remain anchored in the ESA mission definition study report
\citep{2011arXiv1110.3193L}. The critical need for ground-based data is driven by the required photometric redshift accuracy.
The spacecraft brings three infrared bands ($Y,J,H$) and one broad optical band (VIS) of limited use for the photometric
redshifts. A full optical broadband data set such as the LSST $g,r,i,z$ is critical to reach the required photometric
redshifts accuracy across the redshift range explored by \emph{Euclid}. The addition of the LSST $u$-band further improves
the quality of the photometric redshifts, especially at low redshift. The required depths are reported in the same
metric as LSST in  Table~\ref{tab:depth}.\\

\newcommand{\ye}{\cellcolor{yellow}}
\begin{table*}[ht!]
\fontsize{11}{9}
\centering
\caption{\small LSST and \emph{Euclid} Survey Depths}\label{tab:depth}
\setlength\tabcolsep{0pt} 
\newcolumntype{C}[1]{>{\centering}m{#1}} 
\begin{tabular}{llC{1cm}C{1cm}C{1cm}C{1cm}C{1cm}ll}
\rowcolor{yellow!25}\phantom{se} & \emph{ }		& \emph{u}		& \emph{g}		& \emph{r} 	& \emph{i} 	& \emph{z} 	& \emph{Note} & \phantom{se}\\
\rowcolor{blue!20}  \phantom{se} & LSST			& 23.7 	& 24.9 	& 24.4 	& 24.0 	& 23.5 &  Depth reached per single visit  & \phantom{se} \\
\rowcolor{green!20} \phantom{se} & \emph{Euclid}	& 25.4 	& 25.6 	& 25.3 	& 25.3 	& 24.9 &  Average requirement for ground-based complement & \phantom{se} \\
\rowcolor{black!10} \phantom{se} & LSST visits		& 23 	  & 4 	   & 5 	   & 11 	  & 13   &  Number of visits to reach the \emph{Euclid} depths & \phantom{se} \\
\rowcolor{black!10} \phantom{se} & LSST exp. (mn)	& 11.5 	& 2 	   & 2.5 	   & 5.5 	   & 6.5   &Total integration needed (open shutter time) & \phantom{se} \\
\rowcolor{black!10} \phantom{se} & LSST time (mn)	& 15 	& 2.6 	   & 3.3 	   & 7.2	  & 8.5   & Total budget (integration plus overheads) & \phantom{se} \\
\rowcolor{red!15}   \phantom{se} & \multicolumn{7}{l} {\color{black} $\longmapsto$ \emph{We quote the total magnitude for a point source at 5-$\sigma$ with a PSF fitting photometry}} & \phantom{se}\\
\end{tabular}
\end{table*}

We can now answer an important question: how many visits are needed by LSST to fulfill \emph{Euclid} depth requirements?
For all five bands, from $u$ to $z$, 56 LSST
visits are needed, for a total of approximately 28 minutes of integration time, or 37
minutes when considering the overheads (readouts plus slewing).
However,  \emph{Euclid}
will suffer from a zodiacal diffuse light background that varies
from the ecliptic poles to the ecliptic plane by more than 1.2 magnitude. Since \emph{Euclid} will use a fixed integration
time, these space data will not be uniform in depth across the survey footprint and, in consequence, ground-based data should
as much as possible match the space data depth or anchor on its deepest part, which is possible with LSST.

The numbers listed above present the average solution across the entire $15,000$ square degree \emph{Euclid} survey footprint (not all of which could be observed with LSST). Assuming
uniform survey progress for the LSST wide-fast-deep survey, it would take about 10
months to complete the $g$, $r$, $i$, and $z$ observations, and four years for the $u$-band observations.
Therefore, for all bands except $u$, the required depths in
the areas of overlap between \emph{Euclid} and the LSST main survey will be achieved in about the
first year of the LSST survey.  For the south celestial pole, with the current allocation of time,
LSST will not reach the required depths in eight years for  the u band, and will require 4-5 years for the $i$ and $z$-bands, and within 2 years for the $g$ and $r$-passbands.

The {\it Euclid} Wide Survey
will be deeper by \mbox{$\sim 0.3$ magnitudes} near the ecliptic poles, compared to
the average survey depth, due to the dependence on the zodiacal background.
To exploit this extra depth, approximately $1.7$ times more LSST visits would be needed at high ecliptic latitudes compared to the average requirement.
Even deeper LSST photometry would be beneficial for future joint
analyses that could explore the inclusion of galaxies with lower
S/N, e.g. \S\ref{sec:coverage} and Figure\thinspace\ref{fig:zhisto}.



\section{Modes of Coordination}
\label{sec:howto}

In this section, we discuss various modes in which the experiments can coordinate with each other.
These can be broadly separated into several different types of cooperation, based on the types of
dependency that can occur as discussed above, e.g. methodological, calibration, and data exchange.
In addition, within each of these categories, gains can be made that lead to improved/new science,
improved reproducibility, and improved efficiency.

\subsection{Methodological}

There are several ways in which methodological coordination can lead to improved reproducibility and new
science. In developing methodologies for cosmological probes and algorithms to measure these observables,
inter-experiment collaboration is already helping in many ways. As it is difficult to develop
common algorithms to work on the raw-data analysis, catalog-level manipulation to create
summary statistics is  more commonly done. There are many case studies from
current surveys in which common inter-experiment algorithm development leads to more
robust and validated code, for example CAMB \citep{Lewis2000}, GalSim \citep{Rowe2015}, and  SNANA \citep{Kessler2010}.

In addition to individual-experiment cosmological probes, the multi-experiment
combination enables additional statistics to be applied to the joint dataset. Synergies
between experiments will help, for example, by probing different parts of the cosmic expansion
history so that in combination, Stage IV experiments will probe a larger redshift
range than any experiment individually, leading to improved dark energy constraints, e.g. combinations of SN (LSST) and BAO (\emph{Euclid}) distances \citep[see ][]{2015PhRvD..92l3516A}.
By taking the ratio of observables from different experiments, common statistically
limiting effects (e.g. cosmic variance) may be mitigated, a method known as the multitracer
approach \citep{2009PhRvL.102b1302S,2016MNRAS.455.3871A}. Multimessenger (i.e. multiple wavelength) observations of the same area of
sky also allow for additional cosmological probes, e.g. through cluster studies and
kinetic-SZ measurements. Such multimessenger observations also allow for consistency
tests to check for replicability and reproducibility of results, e.g. measuring the weak
lensing power spectrum from two entirely distinct data sets to be consistent \citep[e.g.][]{2016arXiv160701761S}. \emph{Euclid} Deep Fields may also provide extra epochs for the LSST cadenced observations important for transient science as well as proving high-resolution imaging (and grism spectroscopy) of the host galaxies; such data will improve SNe Ia as ``standardizable  candles.”

\subsection{Calibration}

As previously noted, the precise calibration (e.g. photometric, astrometric, and shape) of both surveys will require considerable external information (see \S 2).
This includes astrometric information to establish a coordinate grid on
the sky, galaxy spectra to calibrate photometric redshifts
(\S\ref{sec:photoz}), high-resolution imaging to calibrate measurements
of weak gravitational lensing (\S\ref{sec:shearmeasurement}), and
simulations of what the universe would look like in different
cosmological models. Both surveys will also require a significant effort to achieve their desired photometric accuracy, and coordination on cross-calibrations between the surveys, and with other large-area experiments (e.g. \emph{Gaia}), will be of huge potential benefit to each survey and the whole astronomical community. All of these require an expensive investment of
time and resources (other telescope observations and supercomputers), plus effort to analyze and
reduce the data into usable form. A well-organized effort aimed at the sharing and comparison of calibration data will be needed to reach the full potential of cross-survey calibration.

A key component of both surveys is the use of gravitational weak lensing to measure cosmological parameters. This is at the heart of both \emph{Euclid} and LSST but remains a significant observational challenge given the potential systematic uncertainties and its  low S/N per galaxy. Present cosmological weak lensing measurements have made huge strides in demonstrating the significant promise of this technique but have also found intriguing tensions with other cosmological measurements, which could be new physics or unaccounted systematic errors \citep[e.g.][]{kids,2016arXiv161004606J,2017arXiv170303383H,2017arXiv170700483E,2017arXiv170706627J,2016PhRvD..94b2001A,DESYEAR1}. The comparison of LSST and \emph{Euclid} weak lensing measurements seems to be an excellent risk mitigation given the accuracy demanded by both experiments.
Direct galaxy-by-galaxy comparisons of shapes are likely to be less useful than overall cross-calibration of shear;  the measured `shape' of a galaxy is highly dependent on the method used to measure that shape, and the methods for the space-based \emph{Euclid} data and the ground-based LSST data will differ in significant ways.

\subsection{Data Sharing}

A higher level of coordination can occur when the collaborations agree to exchange or share the actual data that they collect and/or produce. In experiments such as \emph{Euclid} and LSST, where data processing is a rather long chain of actions, this can take many forms that lead to various degrees of scientific benefit and mutual dependencies. Section~\S\ref{sec:cosmology} covered a number of science cases for data sharing between the collaborations. These are explored in more practical terms here, starting from the deepest level of data exchange. Constraints coming from the respective observation schedules of the experiments are not discussed here; they would give rise to a higher level of practical complexity.

 The deepest level of data sharing would occur with pixel-based coordination. At this level, collaborations agree to share access to the lowest stage of their data products. It is likely that, practically speaking, calibration of raw data is still within the remit of each experiment as it requires detailed expertise of instrument scientists, which is not easily shared. Nevertheless, coordination must take place between these instrument scientists if only to define common calibration references. Astrometric reference coordination in the post-\emph{Gaia} era is probably going to be straightforward, but photometric calibration is more demanding, especially since both experiments have extreme requirements on the knowledge of galaxy colors for photometric redshift determination. Thus, to achieve coordination at the pixel level, collaborations should identify a forum to exchange information on standards, instrumental characteristics, and calibration approaches. In such a framework, a very large number of the scientific goals identified in \S~\ref{sec:cosmology} can be reached. A single catalog can be built from detection applied to the complete data set. This catalog then benefits from both the higher spatial resolution of \emph{Euclid} to achieve high-efficiency deblending on the brighter sources and the greater depth of LSST to provide reliability to the fainter sources. In such a configuration, we probably reach the optimal case for photometric redshift determination as well. If  coordination is put in place at the pixel level, then it also allows all the cases of shear measurement calibration/correlation between the two experiments, as the source lists and pixel information are fully compatible.

A slightly less deep level of coordination could be achieved with catalog-only combination. This case is more restricted than the above as here data coordination would occur when both collaborations have created their source catalogs. A driver for this case is to maintain a separation  for each experiment as catalogs are built from two independent data sets. The main purpose of coordinating at the catalog level is to increase (or quantify) the reliability of sources in each catalog. On the LSST side, the high-resolution \emph{Euclid} source list can provide prior information for deblending, while on the \emph{Euclid} side a deeper source list will provide a better control of source reliability at low S/N. Practically, a coordination at catalog level would likely mean two stages of object detection and cataloging, one independent and one using the other catalog for prior information. However, to allow one to combine the catalogs, a number of conditions have to be satisfied; these conditions are similar to those set above. A shared astrometric reference must still  be defined so that cross-matching at the catalog level can be performed. This must be defined early on as astrometric corrections are applied before cataloging takes place. Coordination at the methodological level should also occur to clarify the particular approaches used for object detection and thus significance. Beyond increasing the reliability of sources inside the catalogs, the benefit of coordinating at the catalog level can be to allow the creation of common source lists between the two datasets, for instance, to build samples on which shear measurement calibration can be inspected/compared. A limit of coordination at the catalog level is that it is not sufficient to guarantee the compatibility of photometric measurements, unless explicit coordination has also happened for the definition of a photometric reference system and for the measurement approach itself.

Coordination in photometric measurements is also a possibility. As \S~\ref{sec:cosmology} mentions, both collaborations have an interest in accessing each others' photometric capacities: \emph{Euclid} to cover its wide visible band with a set of narrower filters and LSST to increase its wavelength coverage with the \emph{Euclid} NIR bands. In that context, one can envision a coordination at the photometric measurement level, each collaboration providing, as a service, forced photometry on its imaging data. This is probably the minimum level of coordination (above no coordination at all). It is probably also the level where data independence is maximum. To achieve this, however, once again a number of steps have to be taken so that the imaging datasets are compatible with one another. Quite obviously, these will include defining common astrometric and photometric references, but also sharing information on data processing steps up to the construction of calibrated images (individual and co-added) so that uncertainties in the photometry are properly taken into account.

The above cases thus all show that if coordination between \emph{Euclid} and LSST should 
cover mission data, a number of steps have to be taken whatever the case. For instance, common calibration references should be defined (at least for the astrometric reference) and information should be exchanged on the early stages of data processing, including catalog construction. As both collaborations have already clearly defined processes for data processing, implementing these coordination requirements should be quite feasible.

\subsection{Survey planning}

There are a number of opportunities for coordination within the
planning of the LSST and \emph{Euclid} surveys. These include common survey
footprints, depth as a function of passband, time of observations
(including contemporaneous observations), software for processing and
calibrating the pixel data, common calibration catalogs, and
interfaces for sharing catalogs and resources. Many of these
coordination activities can be accomplished prior to the launch of
\emph{Euclid} or the engineering first light for LSST.

As described in \S\ref{sec:overlap}, for the main LSST survey
area, comprising 7,000 square degrees of overlap, the LSST will reach the
equivalent depth of \emph{Euclid} within the first $\sim$10 months of observations
for all except the $u$-band observations. Prioritizing a uniform
sampling of the sky by the LSST (as opposed to a rolling cadence where
less of the sky is observed but more frequently) and optimizing the
scanning strategy for \emph{Euclid} would maximize the speed at which this
joint dataset could be derived (although the impact of such a change on LSST transient science would need to be studied).
For the $1,000$ square degrees patch of overlap at the South Celestial Pole (SCP), a comparable depth would require between 2 and 8 years of overall LSST operations, dependent on the passband; the SCP is classed as a mini survey within LSST and gets fewer visits than the main survey.
Increasing LSST sky overlap with the \emph{Euclid} survey beyond these two areas would require extending the LSST footprint with a northern $3,000$ square degrees (up to a declination of
+30 degrees) costing  approximately 1\% of the LSST overall survey time.

Early \emph{Euclid} science coordination with the commissioning
activities of the LSST would provide substantial opportunity for the early
calibration of the joint data sets. During commissioning, LSST expects to survey a small
($\sim$\,100 square degree) region to ten-year depth and a wider
($\sim$\,1,000 square degree) region to a two-year depth to validate its source
detection and characterization algorithms.  \emph{Euclid} plans to make regular data releases to the public to promote legacy science as well as stimulate follow-up observations. \emph{Euclid} will have four ``quick'' (Q) release fields (some possibly selected via community input) and three official data releases (DR) spaced regularly throughout the mission with the first quick release $14$ months after the start of survey operations. The quick release products will have undergone Level 2 data processing (co-added images, PSF model and distortion maps, co-added spectra) with the best available calibration at that time, but will be restricted in area ($\simeq50{\rm deg^2}$) and location. The official data releases  will grow in area over the survey lifetime (e.g. 2500, 7500, and 15,000  ${\rm deg^2}$ for DR1, DR2, and DR3 respectively) and also provide Level 2 and Level 3 (e.g. cosmology-related data products) data products. An obvious area of coordination would be to ensure that  LSST observations are available for the \emph{Euclid} quick release fields at the same depth as (or greater than) of the \emph{Euclid} visible and NIR photometry .

Coordination of early
\emph{Euclid} and LSST observations would clearly promote tests of joint
processing algorithms and calibration of the
photometric redshifts and shear measurements. This could occur early on in both surveys, although given the constraints on
the commissioning time for the LSST it is unlikely that any extended-footprint
surveys could be undertaken within that commissioning or early
operations period.

The existence of detailed survey simulation frameworks for both
experiments provides a mechanism by which common commissioning
geometries and the methodologies for synchronization of the survey
strategies (including any advantages from contemporaneous observations) could be evaluated in the
soon. Making such tools widely available would benefit both communities.

In summary, we should aim to exploit as much as possible the complementarity of the two survey strategies, namely with LSST going wide first, building up depth over 10 years, while \emph{Euclid} goes to full depth first, while collecting area over time. The two surveys also have complementary data release philosophies, with both planning to release worldwide significant data sets after the first few years of operations, with additional releases thereafter. Some coordination in these releases would greatly benefit the global community of astronomers, especially in the location of the LSST ``deep drilling fields'' and the  \emph{Euclid} deep fields, as well as the \emph{Euclid} ``quick release'' fields.

\subsection{Data Products and Information Exchange}
\label{sec:dataproducts}
As described above, scientific coordination between \emph{Euclid} and LSST has many aspects and can be pursued to different depths. In order to maintain flexibility and facilitate information exchange, a joint effort might want to design the implementation of joint processing of data from the two missions in tiers of scientific and technical complexity.  The tiers lend themselves naturally to sequential implementation.  This allows different joint processing tasks to be spun off at different tiers, with the more complex tasks addressing more intricate science questions relying on more advanced tiers.

Tier 0 or Data Infrastructure would be aimed primarily at basic comparison of the survey data products for sanity checks.  This tier would start with organizing data for faster access, for instance by coordinating the indexing schemes or regridding the images so pixels align across surveys.  This tier would also include cross-checks on the photometry, astrometry, and depths of surveys to detect any systematic offsets, understand them, and correct for them. This tier would rely primarily on the extraction catalogs from the two surveys to compare astrometry and photometry, but would verify the image product alignments in image space, i.e. in pixel space.  This tier may well be limited to point-source photometry in order to avoid the complications of extended object estimation. The products from Tier 0 would include astrometric distortion maps that represent the differences between the two surveys and photometric offset maps that represent the coefficients needed to convert photometry between the two surveys. The challenge here is to design data structures that capture the departures and systematics without forcing either one of the two data sets to conform to the other.

Tier 1 or Cross-survey Associations would connect individual entries in the catalogs across the two missions, establishing one-to-one as well as one-to-many mappings.  This would require coordination on the databases and their designs. In establishing the mappings, positional and photometric comparisons would be used in the first round, and other parameters would be added if needed in later rounds.  These associations would take into account the effects of varying spatial resolution in the different surveys and the astronomical properties of galaxies, such as their extent on the sky.
They would characterize the severity of confusion for objects or parts of the sky.
The products from Tier 1 would include association tables listing for each object in survey A all of the related objects in survey B, with annotation about the details of the relations, e.g. significance of spatial coincidence, match in brightness, index of local crowding in either survey. Ideally, a single such table would collect all of the information in a symmetric treatment of both surveys.

Tier 2 or Higher-level Checks would address the comparison of more advanced extracted information such as shape measurements or colors of objects.
The products from Tier 2 would include estimation offset maps equivalent to those described in Tier 0 for the photometry.

Tier 3 or Science Analysis would implement pixel-level joint processing, running algorithms that operate simultaneously on images from both surveys, using a variety of algorithms from multi-image estimators to model parameter estimation based on the multiple images and a sky and object model. There are clearly many different branches in this Tier, and they will yield different products, depending on the question being asked.  Obvious examples include blended object disambiguation; photometric/morphological decomposition, with application to photo-z estimation; or using colors to improve shape estimation in very-wide-band images. Tier 3 products could include the products of scientific  analyses and could be in the form of parameters attached to objects in the images or in the form of synthesized images resulting from joint pixel-level processing. Much of the benefit of LSST/\emph{Euclid} coordination  described in this paper will rely on eventual coordination at the Tier 3 level, including the possibility of updating the association tables constructed in Tier 1 with the joint result of higher-level analyses

Lower Tiers (0 and 1) of processing feed into the higher Tiers, and could be implemented infrequently to benefit all subsequent processing, if care is taken to keep the focus on the basic properties of the data rather than specializing to specific science goals. Tiers 0 and 1 products will support all research using both surveys, not just dark energy cosmology.  By contrast, Tiers 2 and 3 will tend to be specialized by scientific question and will need to accommodate many different ways of processing the data.


\section{Computational Issues}
\label{sec:needs}

In this section, we discuss the computational resources and infrastructure that might be needed to perform joint pixel-level processing of \emph{Euclid} and LSST data.

The value of \emph{Euclid} data to LSST includes combining the space-based image data with the LSST images.
The addition of \emph{Euclid} data might be reasonably straightforward for LSST processing,
where it could be added into the processing as if it were a few extra visits, albeit
with higher resolution and significant undersampling. The computational overhead for LSST might be minimal,
with the main issues based on the technical understanding of the \emph{Euclid} data.  This technical understanding would be best enabled by close cooperation of the \emph{Euclid} data processing `Science Ground Segment (SGS)' and the LSST data center(s).

In the standard external data processing paradigm of \emph{Euclid}, the data are first
``Euclidized'' -- converted to a common photometric and astrometric calibration
standard, quality assessed,  weights and masks are generated and converted to a
Euclid-compatible format. This is a standard data combination procedure of building the ``transfer'' function that will make external data compatible with the reference systems that will be used in \emph{Euclid}, and is consistent with the quality requirements for such data. These data then enter the rest of the \emph{Euclid} pipeline.

Processing images that have already been co-added through the LSST pipeline
independently from \emph{Euclid} data is not optimal; however, the benefit of
a full co-processing still needs to be assessed quantitatively.
To realize those benefits, we will need to carry out  multi-image source fitting, where the characteristics of a source are simultaneously fit on all images, even those where the source is below the detection threshold, whereas the source position is imposed from a global astrometric solution with optimized deblending algorithms (see, for instance, \S\ref{sec:objectdetection}).

For \emph{Euclid}, the LSST data could be the largest ingestion of external data, if all (or even many) of the LSST visits are ingested.
To process at the individual image level would require around a factor 50
increase in raw data volume and a factor of around 20 increase in the required
processing power compared to the present DES. This is possible within the context of the
\emph{Euclid} SGS but would require an increase in capacity that would depend on how much of the LSST data were ingested.
However, it is likely that several of the benefits of joint pixel-level processing can already be realized if \emph{Euclid} works with co-added LSST data, and only to the depth of
the \emph{Euclid} data. This imaging depth would be available within the first few years of LSST, which
would reduce the computational requirements for \emph{Euclid} ingesting LSST data considerably.

The Euclidization process could be carried out in two ways. The standard
Euclidization steps could be integrated into the LSST pipeline so that
LSST produced the appropriate calibrated, quality-controlled, weighted, and
masked LSST data for \emph{Euclid}, achieving software reuse on the \emph{Euclid} side, and resource reuse for a
mostly i/o dominated process.
An alternative would be to deploy specific
LSST software within the \emph{Euclid} SGS. Again, this would allow software to be
reused, but potentially require significantly more processing in \emph{Euclid},
unless only stacked data are used;  even then, custom \emph{Euclid} code might be required depending on
LSST's choice of stacking algorithm. Either approach would benefit from  some
synchronization of LSST and \emph{Euclid} processing to control computing costs.

The primary difficulties for LSST in handling \emph{Euclid} data is the undersampling of the NISP images and the
probably correlated noise resulting from resampling as well as the nature of NIR detectors.  Other effects
(e.g. chromatic PSFs, band-dependent morphologies) are already present in LSST images and will be accounted
for by the LSST pipelines.  For VIS data, LSST would ingest individual images with instrumental effects removed
(bias, flatfields, nonlinearity, brighter-fatter, cross-talk, etc.) and would probably also employ the
\emph{Euclid} astrometric and photometric modelling, along with the PSF model.  For NISP data, LSST would ingest
the combined dithered images and possibly also the processed individual images.  In either case, LSST
would employ \emph{Euclid} calibrations including the PSF models --- PSF estimation in undersampled data would
be a significant extension of the LSST data processing and would presumably duplicate work already
performed by \emph{Euclid}.

LSST is considering a variety of algorithms for combining images, and some of the options would not easily
handle the individual NISP exposures; fortunately, hybrid schemes that could accommodate NISP data appear
simple enough to implement.  The LSST image processing algorithms already need to handle a factor of 2.5 in
wavelength (350nm -- 1$\mu m$), and extending this to NIR data may be merely a matter of degree, depending on the difficulties imposed by the undersampling in NISP data.  Additionally,
the LSST data are deep enough that adding \emph{Euclid} data is not expected to add significantly to the number of
objects being characterized.

It is likely that both datasets will need to be ingested into both
pipelines, given that the processing methodologies, driven by the
different scientific use of the data, could well be very different. This approach would also address some of the concerns about loss of independence of the two pipelines and cross-checking of results and cosmological constraints.
A deciding factor here will involve the trade-offs between additional costs
of code adaption and integration, processing, and storage and the
savings from sharing code, storage and processing. How the mutual ingestion of data will be carried out depends on some currently uncertain issues:

\begin{itemize}
 \item Does \emph{Euclid} need to Euclidize all of the LSST data, or if the main
application is colors for photometric redshifts, does it only require a subset
(overlapping on sky and depth, stacked data) of the LSST data?

\item Does the Euclidization step for LSST data most naturally lie
within the \emph{Euclid} or LSST pipelines?
What algorithms will need to be adapted or developed in LSST to ingest
Euclid data, and what will need to be adapted or developed in either pipeline for Euclidization?

\item How much duplication of effort is both desirable and inevitable?

\item What are the difficulties associated with deblending and undersampling?

\end{itemize}

The first issue will drive the subsequent items and so a firm understanding of
what is needed from both surveys needs to be clarified. The size of the LSST
data to be ingested into \emph{Euclid} will also determine the additional storage and processing
needs of the \emph{Euclid} SGS.

One way to address those issues is to create a common \emph{Euclid}/LSST software working group composed of
software and computing experts from both projects and to carry out on a realistic set of tests using simulated data.
This group would then interact with both \emph{Euclid} and LSST software groups in order to adapt the existing pipelines or
develop new ones. Some of the tests should also be designed in order to assess the impact of such a common processing on the
hardware infrastructure. The common \emph{Euclid}/LSST software working group should also work in close contact with both projects' science
working groups in order to design the tests for realistic use cases and to maximize the scientific return.



\section{Conclusions}
\label{sec:conclusions}

We have outlined in this paper many reasons why \emph{Euclid} and LSST may wish to collaborate and coordinate, going well beyond an earlier white paper on dark energy mission coordination \citep{JainEtal2015}
We have outlined the numerous scientific benefits that will come from such coordination in pixel-level processing, cadence and survey overlap, calibration data acquisition, and resource allocation (especially in high-performance computing). The benefits we explored are primarily in dark energy cosmology, but we have touched on other areas of scientific benefit, including galaxy evolution, dark matter studies, and solar system science. We have offered a plausible path forward that will require further scientific development by the community to realize, as well as coordinated discussions between the two projects and the  relevant funding agencies. The two projects will have to come to some agreement on the political issues surrounding data access to enable the benefits outlined in this paper. We feel that this path will lead to the maximum scientific return from \emph{Euclid} and LSST in the 2020s and into the 2030s.

\textbf{Acknowledgments:} J.R. and A.K. were supported by JPL, which is run under a  contract for NASA by Caltech.  J.R., A.K., P.C., R.B., R.L. and R. Mandelbaum were supported in part  by NASA ROSES  grant 12-EUCLID12-0004. A.N.T. is supported by a UK Space Agency {\em Euclid} grant and a Royal Society Wolfson Research Merit Award. Y.M. was supported by the French CNES Space Agency and Institut National des
Sciences de l'Univers of CNRS. T.S. acknowledges travel support from the German Federal
Ministry for Economic Affairs and Energy (BMWi) provided
via DLR under project No. 50QE1103. T.K. is supported by a Royal Society University Research Fellowship. V.F.C. is funded by the Italian Space Agency (ASI) through contract Euclid - IC (I/031/10/0) and acknowledges financial contribution from the agreement ASI/INAF/I/023/12/0. R. Massey acknowledges support from UK STFC grant ST/N001494/1. B. N. is supported by a UK Space Agency Euclid grant.
 This paper benefitted from ideas developed in the UK Dark Energy Strategy 2020 Document\footnote{\url{https://www.ucl.ac.uk/mssl/astro/Documents/UK_Dark_energy_strategy_2020/}}.
We thank Michael Brown, Elisa Chisari, Henk Hoekstra, \v{Z}eljko Ivezi{\'c},  Bhuvnesh Jain, Roberto Scaramella, and Peter Schneider  for useful feedback on this paper. We thank Katrin Heitmann for useful discussions on high-performance computing. We thank the anonymous referee for useful suggestions on adding quantitative rigor to sections of this paper and useful feedback on the structure of the paper.

\appendix
\section{Appendix: Thoughts on Coordination}
\label{sec:appendix}
\subsection{Interdependency and Independence}
\label{sec:independence}

A key question is what level of cooperation is appropriate at each stage of the development of these experiments.
One extreme scenario is that the two experiments work in complete isolation and
simply exchange final results, already fully analyzed independently, for combined scientific constraints. Many recent breakthroughs in physics were carried out this way;
for example, CMS and ATLAS \citep{2012PhLB..716....1A, 2012PhLB..716...30C}, 2dFGRS and SDSS \citep{2005MNRAS.362..505C, 2005ApJ...633..560E}, and the supernova dark energy experiments of \citet{1998AJ....116.1009R} and \citet{1999ApJ...517..565P}.

However, before combining the results of two experiments, the
consistency between the individual
results must be established. Such analyses may not have been ``blinded'' (see below), thus adding more possible uncertainty in such a comparison. In the event that an inconsistency arises between the experiments, then a joint study must be done to disentangle the reasons for this inconsistency. Such tests, after the fact, then rely on the full coordination of both teams, who may have changed over the many years of these long-term projects.

At the other extreme, the two experiments could merge all parts of their data
processing, thus entirely removing independence and creating a de facto single
experiment.

Between these extremes lie a broad spectrum of possibilities
that balance independence against potential gains of closer cooperation.
Some potential gains are:
\begin{itemize}
\item
The combination of data with different unknown biases can lead to improved science through the reduction of systematic
uncertainties in the measured parameters. Likewise, coordination in survey overlap and cadence can improve science output.
\item
New science can be enabled through the combination of data and through
cross-correlation statistics that are not possible with a single experiment alone.
\item
An increase in the reproducibility of results. If two experiments share
data analysis stages, then potential systematic effects arising from implementation
differences between algorithms (for example) are mitigated.
%
\item
The additional data from each experiment can enable (blind) validation tests which could not
be carried out in isolation and without access to both datasets.
\item
There can be efficiency savings in terms of sharing resources, particularly
data or computationally intensive tasks.
\end{itemize}

The potential cost of cooperation between experiments is primarily a loss of independence.
Independence is a critical aspect of the scientific methodology and underpins our confidence in  our body of knowledge.
If a Stage IV experiment discovers that our Standard Model of cosmology is incomplete, or uncovers some deviation from general relativity, then
external replication of this result will be critical for
scientific credibility.

Any consideration of cooperation between experiments is therefore a trade-off between
 replicatability versus reproducibility, plus gains in new and improved science
and efficiency.

Although it is only possible to independently replicate a
measurement via an independent experiment, it is first necessary to clearly define
``dependency'' and ``independency''.
In this context, dependency between experiments is primarily gained via three routes. First, the experiments may analyze their data using the same methods, algorithms, and codes. Second, the experiments may use the same external calibration data or numerical simulations to test or train data analysis techniques. Finally, raw data products from the two experiments may be analyzed simultaneously (for example,
in the expectation that information
from one experiment may help to reduce systematic errors in the other).

For the specific case of LSST and \emph{Euclid},
it is crucial to recognize that some interdependency between experiments is inevitable,
by virtue of the sociology and size of the cosmology community, and the restricted set
of available data.
The two experiments already have a large (and growing) overlap in personnel and have been shaped by some of
the same people. Arranging non-overlapping teams would now be
impossible without major restructuring of consortia\footnote{At the time of this writing, the \emph{Euclid} Consortium has over 1400 members and the LSST Dark Energy Science Collaboration has over 600.}.
Moreover, \emph{Euclid} and LSST will observe the same sky -- there is only one universe to observe --
because the survey areas are large enough that overlap is inevitable.

Shared cosmological simulations are an obvious place for cost savings. Also,
they will have to use similar external calibration data. For instance, only the \emph{Hubble Space Telescope (HST)} has the capability of observing higher-quality data over a smaller area (prior to the launch of \emph{WFIRST}), and
both surveys will  use \emph{Gaia} \citep{gaia} data for
stellar SEDs and
astrometric and photometric information. Both surveys will depend on
deep spectroscopic data sets for photo-z calibration \citep[e.g.][]{2015ApJ...813...53M}.
Therefore, traditional methods of running separate experiments on different underlying data are not possible.

Nonetheless, we argue that the design characteristics of LSST and \emph{Euclid} mean that
coordination can be managed in a way that not only avoids some of the downsides of dependency,
but addresses the uncontrolled dependencies that have emerged naturally.
\begin{enumerate}
\item
The two experiments will use different instrumentation and observe from different environments and with different strategies:
LSST from the ground at high cadence, and \emph{Euclid} from space at high resolution.
Therefore, even though personnel are overlapping,
the analysis of raw data will necessarily require different algorithms.
\item
Within each experiment, independence is already maintained between cosmological probes.
For example, the methods,
algorithms, and calibration data required for weak lensing, galaxy clustering, and supernovae studies are
to a  large degree mutually exclusive.
\item
\emph{Euclid} and LSST have already established separate data analysis pipelines, built on quite different platforms and philosophies. Therefore, combined data analysis will be performed
in addition to, not instead of, separate analyses of both data sets.
\end{enumerate}

Furthermore, a commonly agreed strategy between the experiments on blinding methods, implementing
validation tests, and open-source publishing of algorithms and raw data will
lead to verifiable results. This may be challenging given the different cultures of the two experiments, but this argues for closer immediate cooperation to discuss and agree upon such procedures.

We must also address confirmation bias in cosmology, namely, that results tend to confirm investigators' prior
beliefs through unconscious selection of data and results. This is
not merely a theoretical supposition, but has been observed in the literature \citep{2011arXiv1112.3108C}.
However, the perception of the extent of this problem is somewhat subjective,
and scientists are arguably more immune to such psychological effects than most, because they are trained to be skeptical and question results even if they appear to agree.
Blind analysis may be able to mitigate against this type of confirmation bias \citep[e.g.][]{kids,2016PhRvD..94b2001A,2017MNRAS.465.4895W,2015MNRAS.454.3500K,2017MNRAS.471.2254Z}. These serious issues of scientific credibility
and veracity of results are areas of common concern to experiments and therefore
common cooperation on solutions is desirable.

Within the context of veracity of cosmological results, one important aspect
of coordination is that of risk mitigation. Through coordination, various risks can be mitigated
that could threaten to undermine the cosmological results from either experiment alone, e.g., if the two surveys yield inconsistent results, a natural step would be to
compare the individual measurements in detail, to find out where, and why, they disagree. A risk in this scenario is that lack of coordination will lead to a protracted period of comparison
during which the perception of the robustness of both results is undermined. This comparison could become acrimonious, increasing the possibility that neither team would want to appear to have the ``wrong'' answer.

This risk is mitigated by anticipating this scenario and planning for comparison steps that
have been coordinated from the outset, e.g. by sharing a common coordinate system, defining a common patch of the sky, and/or
even a common source detection catalog. The minimum level of cooperation required is therefore one of
scenario planning and developing risk-mitigation strategies, as well as developing the political desire to promote such coordination.

\subsection{What if We Do Not coordinate}

If the communities behind LSST and \emph{Euclid }do not engage in an organized effort to conduct joint processing, the world will not stand still. There will be many small efforts attempting to fill this obvious gap, and they will meet with various degrees of success. The certain outcome of this situation will be the wasted effort resulting from the many independent attempts to recreate the same basic information (Tiers 0 and 1 as described in \S\ref{sec:dataproducts} at least, and definitely more).  To the further detriment of the field, the lack of such a collaborative effort will limit the exchange of deep expertise about the surveys, so that each independent effort at attempting the joint processing will fall short in its understanding of either one or the other data set, resulting in slower progress, distorted approaches, and potentially to wrong conclusions. Ultimately, the outcome will be less science on both sides, and less science from the combined data.  Given the pressure to march on to the next stage of dark energy experimentation, this opportunity of joint processing may be lost forever, along with potential insights that would have informed the design of the next stage.

\subsection{Culture}

Both LSST and \emph{Euclid} are large, international collaborations with many common managerial issues (e.g. publication policies, membership and reward tracking, meetings, etc.). Both experiments have developed different cultures partly due to their origins. We would hope that some sharing of good practice and experience in these cultural issues would be beneficial for both experiments.

Moreover, both experiments will have a long-term need for well-trained people to help analyze and understand the data and results. Coordination in the training of such early career researchers would be beneficial.  Both projects will benefit from interchanges and joint training of future leaders in  the advanced statistical and computational techniques required (i.e. data scientists). Such skills are also required for the global economy, and astronomy provides an excellent training environment.

\bibliographystyle{apj}
\bibliography{references}

\end{document}